\documentclass[11pt,fleqn]{article}

\usepackage{amsmath,amssymb,amsthm,enumerate,cite}%,backref}%

\newcommand{\p}{\partial}

\newcommand{\const}{\mathop{\rm const}\nolimits}
\newcommand{\sech}{\mathop{\rm sech}\nolimits}
\newcommand{\csch}{\mathop{\rm csch}\nolimits}
\newcommand{\sn}{\mathop{\rm sn}\nolimits}

\newcommand{\vspacebefore}{\raisebox{0ex}[2.5ex][0ex]{\null}}

\newcounter{tbn}

\newcounter{tabu}

{\theoremstyle{definition}

\newtheorem{note}{Note}
\newtheorem*{note*}{Note}
\newtheorem*{example*}{Example}
\newtheorem{example}{Example}

}

\begin{document}

\par\noindent
{\LARGE\bf Algorithmic framework for group analysis\\ of differential equations and its application\\ to generalized Zakharov--Kuznetsov equations

 %Lie group analysis of differential equations\\ illustrated by example\\ of generalized Zakharov--Kuznetsov equations
 \par}

{\vspace{4mm}\par\noindent {\bf Ding-jiang Huang$^\dag$ and Nataliya M. Ivanova$^\ddag{}$
} \par\vspace{2mm}\par}

{\vspace{2mm}\par\noindent {\it
$^\dag$~Department of Mathematics, East China University of Science and Technology,  Shanghai,\\
$\phantom{^\dag}$~200237, P. R. China\\
}} {\noindent \vspace{2mm}{\it $\phantom{^\dag}$~e-mail:
djhuang.math@gmail.com, djhuang@ecust.edu.cn}\par}

{\vspace{2mm}\par\noindent {\it
$^{\ddag}$~Institute of Mathematics of NAS of Ukraine, %\\
%$\phantom{^\dag$~
3 Tereshchenkivska Str., 01601 Kyiv, Ukraine\\
}}
{\noindent \vspace{2mm}{\it $\phantom{^\ddag}$~e-mail: ivanova@imath.kiev.ua }\par}

{\vspace{5mm}\par\noindent\hspace*{8mm}\parbox{145mm}{\small
In this paper, we explain in more details the modern treatment of the problem of group classification of (systems of) partial differential equations (PDEs) from the algorithmic point of view.
More precisely, we revise the classical Lie--Ovsiannikov algorithm of construction of symmetries of differential equations,
describe the group classification algorithm and discuss the process of reduction of (systems of) PDEs to (systems of) equations with smaller number of independent variables in order to construct invariant solutions.
The group classification algorithm and reduction process are illustrated by the example of the generalized Zakharov--Kuznetsov (GZK) equations of form
$u_t+(F(u))_{xxx}+(G(u))_{xyy}+(H(u))_x=0$. As a result, a complete group
classification of the GZK equations is performed and a
number of new interesting nonlinear invariant models which have
non-trivial invariance algebras are obtained.
%It is shown that the symmetry algebra of invariant models from the GZK equations, is at most six-dimensional.
Lie symmetry reductions and exact solutions for two important invariant models, i.e., the classical and modified Zakharov--Kuznetsov equations, are constructed. The algorithmic framework for group analysis of differential equations presented in this paper can also be applied
to other nonlinear PDEs.
%35L65, 35Q60, 70H07, 70H33%, 58J72 22E05, 22E60, 35C05(Solutions in closed form), 35A22 Transform methods (e.g. integral transforms),
%35A30 Geometric theory, characteristics, transformations,58J70 Invariance and symmetry properties [See also 35A30],58J72 Correspondences and other transformation methods (e.g. Lie-Backlund) [See also 35A22],35K55 Nonlinear parabolic equations, 35Q60 PDEs in connection with optics and electromagnetic theory

}\par\vspace{5mm}}

\section{Introduction}

One of the most famous two-dimensional generalizations (together with Kadomtsev-Petviashvili equation) of the
Korteweg-de Vries (KdV) equation is given by the Zakharov--Kuznetsov (ZK) equation
\begin{equation}\label{eq2DimZKEq}
u_t+au_{xxx}+bu_{xyy}+cuu_x=0.
\end{equation}
It was first
derived by Zakharov and Kuznetsov~\cite{Zakharov&Kuznetsov1974} to describe nonlinear ion-acoustic waves in a magnetized plasma.
More precisely, they considered a plasma in a strong magnetic field, {\mathversion{bold}${\bf B}={ B\hat z}$}, with
cold ions and hot electrons ($T_e\gg T_i$). The ion motions are
described by the following equations
\begin{gather*}
n_t+\nabla\cdot(n{\bf u})=0,\quad {\bf u}_t+({\bf u}\cdot\nabla){\bf u}=-\frac e{m_i}\nabla\phi+{\bf u}\times{\bf \Omega}_i,\\
\nabla^2\phi=-4\pi e(n-n_e),
\end{gather*}
where $n$ is number density of ions,
${\bf u}= (u,v, w)$ is ion velocity,
$m_i$ is ion mass,
$\phi$ is electric potential
${\bf \Omega}_i=\frac{e{\bf B}}{m_ic}$ is scaled magnetic field and
$n_e =\exp (\frac{e\phi}{KT_e})$.

After introducing the dimensionless variables and approximating the $x$-component
of ${\bf u}$ by the polarization shift these equations look like
\begin{gather*}
n_t-(n\phi_{tx})_x+(nw)_z=0,\quad
w_t=\phi_{tx}w_x+ww_z-\phi_z,\\
\alpha\phi_{xx}+\phi_{zz}=e^\phi-n.
\end{gather*}
Now, after a change of independent variables
$\xi=\varepsilon^{1/2}(z-t)$, $\eta=\varepsilon^{1/2}x$, $\tau=\varepsilon^{3/2}t$
assuming a solution of the latter equations of the form
\[
n=1+\sum_{j=1}^\infty \varepsilon^jn_j,\quad
\phi=\sum_{j=1}^\infty \varepsilon^j\phi_j,\quad
w=\sum_{j=1}^\infty \varepsilon^jw_j,
\]
one gets that $n_1 =\phi_1 = w_1$ with $\phi_1$ being solution of
\[
\phi_{1,t}+\phi_1\phi_{1,\xi}+\frac12(\phi_{1,\xi\xi\xi}+(1+\alpha)\phi_{1,\xi\eta\eta})=0
\]
which has the form~\eqref{eq2DimZKEq}.

In the more realistic situation in which
the electrons are non-isothermal, Munro and Parkes~\cite{Munro&Parkes1999,Munro&Parkes2000}  showed that, with an
appropriate modified form of the electron number density proposed by Schamel~\cite{Schamel1972},
a reductive perturbation procedure leads to a modified form of the Zakharov--Kuznetsov (mZK1)
equation, namely
\[
u_t+au^{1/2}u_x+bu_{xxx}+cu_{xyy}=0.
\]

One more modification of the Zakharov--Kuznetsov equation is given by
Kakutani and Ono~\cite{Kakutani&Ono1965} who have shown that the modified KdV equation governs
the propagation of Alfv\'en waves at a critical angle to the undisturbed
magnetic field. The presence of the transverse dispersion has been physically
attributed to the finite Larmor radius effects~\cite{Hasegawa&Uberoi1982}.
The resulting two-dimensional equation in this physical context is known as the modified
Zakharov--Kuznetsov equation (mZK2)~\cite{Blaha&Laedke&Spatschek1989}
\[
u_t + au^2u_x + u_{xxx} + u_{xyy} = 0.
\]
Here, the sign of $a$ can not be forced to be definite by scaling
considerations and the two signs of $a$ correspond to different physical phenomenon. For example, the
focusing equation can be derived as a model for the evolution of ion
acoustic perturbations with a negative ion component, while the defocusing equation
models the evolution of ion acoustic perturbations in a plasma with two
negative ion components of different temperature.

In order to encompass as many physical applications as possible,
many researchers consider dispersive models of the Zakharov--Kuznetsov (dZK) type
equations of form \cite{Wazwaz2004}
\[
u_t+a(u^n)_{xxx}+b(u^m)_{xyy}+c(u^k)_x=0,
\]
or even the generalized Zakharov--Kuznetsov (GZK) equations
\begin{equation}\label{eq2DimDispZKEq}
u_t+(F(u))_{xxx}+(G(u))_{xyy}+(H(u))_x=0
\end{equation}
with enough smooth functions $F(u)$, $G(u)$ and $H(u)$.

Many mathematical properties such as the stability or transverse instability of solitary-wave solutions, initial-boundary value
problems, generalized Painlev\'{e} formulation, compactons and solitons and so on for the special cases of class~\eqref{eq2DimDispZKEq} have been
investigated exhaustively by many authors~\cite{Blaha&Laedke&Spatschek1989,Johnson2009,Kakutani&Ono1965,Munro&Parkes1999,
Munro&Parkes2000,Wazwaz2004,Shivamoggi&Rollins1991,Shivamoggi&Rollins&Fanjul1993,Qu&Zhu2009,Bridges&Reich2001,Iwasaki&Toh&Kawahara1990,
Chen2003,Sipcic&Benney2000,Faminskii2008,Batiha2008,Hu2008}. However, despite of great interest of researchers and importance of class~\eqref{eq2DimDispZKEq}, very few facts of its Lie symmetry structure and related topics are known.
Therefore, for the sake of providing more information to understand the mathematical structures of the ZK-like equations, in this paper we will perform detailed group analysis for the class of GZK equations~\eqref{eq2DimDispZKEq}, where $F(u)$, $G(u)$ and $H(u)$ are arbitrary smooth (analytic) functions, $F_uG_uH_{uu}\ne0$.

It is known that the Lie group analysis is a systematic and powerful method for handling partial
differential equations (PDEs) \cite{Ovsiannikov1982,Olver1986,Bluman&Anco2002,Galaktionov&Svirshchevskii2006}.
Moreover, it forms a basis for many useful techniques in
both pure and applied areas of mathematics, physics, and mechanics, etc.
For the PDEs, admitting symmetry is an essential part of their intrinsic nature. Based on the symmetries of a PDE, one can successively consider many
other important properties of the equation such as integrability, conservation laws and linearizations, reducing
equations and invariant solutions, fundamental solution and invariant numerical integrators and so on \cite{Ovsiannikov1982,Olver1986,Bluman&Anco2002,Bihlo&Popovych2012,Pang&Temuerchaolu2010,Galaktionov&Svirshchevskii2006,Sinkala&Leach2008,Craddock&Lennox2007,
Liu&Geng2013,Hairer&Lubich&Wanner2002}. In general, for a single PDE one can directly implement
the classical Lie--Ovsiannikov method to compute the symmetries. However, for parameterized classes of PDEs
(namely, equations containing arbitrary constants or functions), one can not derive all the symmetries by direct usage of this method.
One will face the so called group classification problem of PDEs,
which is the keystone of group analysis of differential equations.
Although this problem has been widely investigated for different subclasses of~\eqref{eq2DimDispZKEq}, many of the existing classifications are incomplete.
In fact, one can find a huge number of recent papers on symmetry analysis of PDEs,
including different generalizations of the Zakharov--Kuznetsov equations, where the group classification problem is solved incompletely or incorrectly and
papers studying symmetries of some fixed equations with low physical motivation.
There are also many papers on ``preliminary group classification'' where authors list some cases with new symmetries but do not claim
that the general classification problem is solved. In many respects this can be explained to two main reasons:
(i) many researchers do not incorporate the equivalence transformation theory to the classification problem;
(ii) overdetermined systems of PDEs derived from the invariance criterion of parameterized PDEs under consideration often can not be solved completely.

That is why, in this paper we will describe an algorithmic framework for group classification of (systems of) partial differential equations. More precisely, we revise the classical Lie--Ovsiannikov method of construction of symmetries of differential equations in more details and write down the precise formulation of the algorithm of group classification of a class (of systems) of differential equations by introducing equivalence transformation theory and a compatibility method for solving the overdetermined system of PDEs.
Moreover, we describe a systematic way of reduction of (systems of) partial differential equations to (systems of) equations
with smaller number of independent variables so that we can find all possible invariant solutions of (systems of) differential equations.
We will illustrate this well-known theoretical background and algorithmic framework by the running example of the GZK equations \eqref{eq2DimDispZKEq}.

Therefore, the purpose of this paper is two fold.
On the one hand, we explain the modern treatment of the problem of group classification of (systems of) PDEs from the algorithmic point of view.
On the other hand, we perform the systematic complete group classification and construct invariant reductions for the GZK equations \eqref{eq2DimDispZKEq}. The rest of this paper is organized as follows.
In section~\ref{SectionPrelimi}, we describe the algorithmic framework for group classification of (systems of) PDEs
and give an exhaustive algorithm of solving such problems. An efficient algorithm of constructing optimal systems
of subalgebras of Lie symmetry algebras and invariant solutions of differential equations is also given. In section~\ref{SectionEquivTrasf}, we investigate
the equivalence transformations of the GZK equations  \eqref{eq2DimDispZKEq}. The complete group classification of class
\eqref{eq2DimDispZKEq} is presented in section~\ref{SectionLieSymGZKEq} by using a compatibility method. Section~\ref{SectionExactSol} contains results on optimal systems of
subalgebras of Lie symmetry algebras of two equations from class~\eqref{eq2DimDispZKEq}.
Invariant solutions of the equations under consideration are also constructed.
Finally, some conclusion and
discussion are given in section~\ref{SectionOnConclusion}.

\section{Algorithmic framework for group analysis of differential equations}\label{SectionPrelimi}

\subsection{Computation of Lie symmetries of differential equations}

For construction of symmetries of differential equations we use Lie--Ovsiannikov infinitesimal criterion of invariance~\cite{Ovsiannikov1982,Olver1986}.

Consider the system~$\mathcal L$: $L(x,u_{(p)})=0$
of $l$~differential equations
for $m$~unknown functions $u=(u^1,\ldots,u^m)$
of $n$~independent variables $x=(x_1,\ldots,x_n)$.
Here $u_{(p)}$ denotes the set of all the derivatives of~$u$ with respect to $x$
of order not greater than~$p$, including $u$ as the derivatives of the zero order.
$L=(L^1,\ldots,L^l)$ is a tuple of $l$ fixed functions depending on $x,$ and $u_{(p)}$.

Let \raisebox{0ex}[0ex][0ex]{$\mathcal L_{(k)}$} denote the set of all algebraically independent differential consequences
 of the system~$\mathcal L$ that have, as differential equations, orders not greater than $k$.
Under the local approach, the system~\raisebox{0ex}[0ex][0ex]{$\mathcal L_{(k)}$} is identified with the manifold
determined by~\raisebox{0ex}[0ex][0ex]{$\mathcal L_{(k)}$} in the jet space~$J^{(k)}$.

Each one-parametric group of point transformations
that leaves the system~$\mathcal L$ invariant corresponds to an
infinitesimal symmetry operator of the form
\[
Q=\xi^i(x,u)\p_{x_i}+\eta^a(x,u)\p_{u^a}.
\]
Here and below the summation over the repeated indices is assumed.
The indices~$i$ and~$a$ run from~1 to~$n$ and from~1 to~$m$, respectively.

The infinitesimal criterion of invariance of the system~$\mathcal L$
with respect to the Lie symmetry operator~$Q$ has the form
\[
Q_{(p)} L(x,u_{(p)})\big|_{\mathcal L_{(p)}}=0, \quad\mbox{where}\quad
Q_{(p)}:=Q+\sum_{0<|\alpha|{}\leqslant  p} \eta^{a\alpha}\p_{u^a_\alpha},
\]
i.e., the result of acting by $Q_{(p)}$ on $L$ vanishes on the manifold~$\mathcal L_{(p)}$.
Here $Q_{(p)}$ denotes the standard $p$-th prolongation of the operator~$Q$,
coefficient $\eta^{a\alpha}=D_1^{\alpha_1}\ldots D_n^{\alpha_n}Q[u^a]+\xi^iu^a_{\alpha,i}$,
operator $ D_i=\p_i+u^a_{\alpha,i}\p_{u^a_\alpha}$
is the operator of total differentiation with respect to the variable~$x_i$,
and $Q[u^a]=\eta^a(x,u)-\xi^i(x,u)u^a_i$ is the characteristic of operator $Q$, associated with~$u^a$.
The tuple $\alpha=(\alpha_1,\ldots,\alpha_n)$ is a multi-index,
$\alpha_i\in\mathbb{N}\cup\{0\}$, $|\alpha|\mbox{:}=\alpha_1+\cdots+\alpha_n$.
The variables $u^a_\alpha$ and $u^a_{\alpha,i}$ of the jet space $J^{(r)}$ correspond to the derivatives
\[
\frac{\p^{|\alpha|}u^a}{\p x_1^{\alpha_1}\ldots\p x_n^{\alpha_n}}
\quad\mbox{and}\quad
\frac{\p^{|\alpha|+1}u^a}
{\p x_1^{\alpha_1}\ldots\p x_{i-1}^{\alpha_{i-1}}\p x_i^{\alpha_i+1}\p x_{i+1}^{\alpha_{i+1}}\ldots\p x_n^{\alpha_n}}
\quad\mbox{respectively}.
\]

\begin{example}
For equations of form~\eqref{eq2DimDispZKEq} we look for the infinitesimal symmetry generator in form
\[
Q=\tau(t,x,y,u)\p_t+\xi(t,x,y,u)\p_x+\zeta(t,x,y,u)\p_y+\eta(t,x,y,u)\p_u.
\]
(In the above notation $(t,x,y)=(x_0,x_1,x_2)$ and $\tau=\xi^0$, $\xi=\xi^1$, $\zeta=\xi^2$.)
Application of the Lie infinitesimal criterion to~\eqref{eq2DimDispZKEq} gives
%Ut+F[U, U, U]*Ux^3+3*F[U, U]*Ux*Uxx+F[U]*Uxxx+G[U, U, U]*Ux*Uy^2 +2*G[U, U]*Uy*Uxy+G[U, U]*Ux*Uyy  +G[U]*Uxyy+H[U]*Ux = 0
\begin{gather*}
\eta^t+\eta F_{uuuu}u_x^3+3\eta^xF_{uuu}u_x^2+3\eta F_{uuu}u_xu_{xx}+3F_{uu}(\eta^xu_{xx}+\eta^{xx}u_x)+\eta F_{uu}u_{xxx}\\
+\eta^{xxx}F_u+\eta G_{uuuu}u_xu_{y}^2+G_{uu}(\eta^xu_y^2+2\eta^yu_xu_y)+2\eta G_{uuu}u_yu_{xy}\\
+2G_{uu}(\eta^yu_{xy}+\eta^{xy}u_y)+\eta G_{uuu}u_xu_{yy}+G_{uu}(\eta^x\eta_{yy}+\eta^{yy}u_x)\\
+\eta G_{uu}u_{xyy}+\eta^{xyy}G_u+\eta H_{uu}u_x+H_u\eta^x=0.
\end{gather*}
One can also verify that the coefficients of the first prolongation of $Q$ look like
\begin{gather*}
\eta^t=\eta_t+\eta_uu_t-u_t(\tau_t+\tau_uu_t)-u_x(\xi_t+\xi_uu_t)-u_y(\zeta_t+\zeta_uu_t) ,\quad\\
\eta^x=\eta_x+\eta_uu_x-u_t(\tau_x+\tau_uu_x)-u_x(\xi_x+\xi_uu_x)-u_y(\zeta_x+\zeta_uu_x) ,\quad\\
\eta^y=\eta_y+\eta_uu_y-u_t(\tau_y+\tau_uu_y)-u_x(\xi_y+\xi_uu_y)-u_y(\zeta_y+\zeta_uu_y).
\end{gather*}
In an analogous way the higher order coefficients can be found.
\end{example}

%\begin{example}
%Lie symmetries of the ZKeq? mZKeq? other fixed eq?
%\end{example}
%
%*****************************************************************************

\begin{example}
Consider the Lie symmetries of the ZK equation:
\[
u_t+au_{xxx}+bu_{xyy}+cuu_x=0.
\]
Applying the Lie infinitesimal criterion to the above equation we get
\[
\eta^t + a\eta^{xxx} + b \eta^{xyy} + c\eta u_x + c u \eta^x =0.
\]
Substituting the coefficients of the prolongation of the operator $Q$ into the above equation and splitting it with respect to the unconstrained variables,
we obtain the system of determining equations
\begin{gather*}
\tau_t = 3 \xi_x, \quad \tau_x = 0, \quad  \tau_y = 0, \quad  \tau_u =0; \\
\xi_t = c \eta + 2 c u \xi_x, \quad \xi_{xx} = 0, \quad   \xi_y = 0, \quad  \xi_u =0; \\
\zeta_t = 0, \quad    \zeta_x = 0, \quad \zeta_y = \xi_x, \quad \zeta_u =0; \\
\eta_{uu} = 0, \quad  \eta_t + a \eta_{xxx} + b \eta_{xyy} + c u \eta_x = 0.
\end{gather*}
The general solution of this system is
\[
\tau = c_1+c_2t,\quad \xi = \frac{1}{3}c_2x+c_3t+c_4,\quad \zeta = \frac{1}{3}c_2y+c_5,\quad \eta = -\frac{2}{3}c_2u+\frac{1}{c}c_3,
\]
where $c_1,\ldots,c_5$ are arbitrary constants.
Therefore the most general infinitesimal symmetry generator  has the form
\[
Q= (c_1+c_2t) \p_t + \Big(\frac{1}{3}c_2x+c_3t+c_4\Big)\p_x+\Big(\frac{1}{3}c_2y+c_5\Big) \p_y + \Big(-\frac{2}{3}c_2u+\frac{1}{c}c_3\Big) \p_u
\]
As $c_1,\ldots,c_5$ are arbitrary constants, we conclude that the maximal Lie invariance algebra of the equation under consideration is 5-dimensional
and is spanned by the following generators
\[
\langle \p_t, \  \p_x, \  \p_y,\  3t\p_t+x\p_x+y\p_y-2u\p_u,\
ct\p_x+\p_u \rangle.
\]
\end{example}

\begin{example}
Consider the Lie symmetries of the the modified ZK1 equation:
\[
u_t+au^{\frac12}u_x+bu_{xxx}+cu_{xyy}=0.
\]
Application of the Lie infinitesimal criterion to the above equation gives
\[
\eta^t + \dfrac12 a \eta u^{-\frac12} u_x +  a \eta^x u^{\frac12} + b \eta^{xxx} + c \eta^{xyy}  = 0,
\]
or, multiplying it by two for convenience,
\[
2\eta^t u^{\frac12}  +  a \eta  u_x + 2 a \eta^x u + 2 b \eta^{xxx} u^{\frac12}  + 2 c \eta^{xyy}  u^{\frac12}  = 0.
\]
Substituting the coefficients of the prolongation of the operator $Q$ into the above equation, we obtain the system of determining equations
\begin{gather*}
\tau_x = 0, \quad \tau_y = 0, \quad \tau_u = 0,\quad \tau_t = 3 \xi_x, \quad \tau_{tt}=0;  \\
\xi_y = 0, \quad \xi_u = 0, \quad \xi_{xx} = 0, \quad \xi_{xt} = 0; \\
\zeta_t = 0, \quad \zeta_x = 0, \quad \zeta_u = 0, \quad \zeta_y = \xi_x;\\
\eta_{xu} =0, \quad \eta_{yu} =0, \quad \eta_{uu} =0, \quad a \eta - 2 u^{\frac12} \xi_t  + 4 a u \xi_x  = 0,\\
\eta_t  + a u^{\frac12} \eta_x + b \eta_{xxx} + c \eta_{xyy} = 0.
\end{gather*}
Its general solution is
\[
\tau = c_1+c_2t,\quad \xi = \frac{1}{3}c_2x+c_3,\quad \zeta = \frac{1}{3}c_2y+c_4,\quad \eta = -\frac{4}{3}c_2u,
\]
where $c_1,\ldots,c_4$ are arbitrary constants.
Therefore the most general infinitesimal symmetry operator has the form
\[
Q= (c_1+c_2t) \p_t + \Big(\frac{1}{3}c_2x+c_3\Big)\p_x+\Big(\frac{1}{3}c_2y+c_4\Big) \p_y -\frac{4}{3}c_2u \p_u,
\]
that implies that the Lie algebra of infinitesimal symmetry generators can be represented as
\[
\langle \p t, \ \p x, \ \p y,\ 3t\p t+x\p x+y\p y-4u\p u \rangle.
\]
\end{example}

\begin{example}
Consider the Lie symmetries of the the modified ZK2 equation:
\[
u_t + au^2u_x + u_{xxx} + u_{xyy} =0
\]
Application of the Lie infinitesimal criterion to the above equation gives
\[
\eta^t + \eta^{xxx} + g\eta^{xyy} + 2h\eta u u_x + h \eta^x u^2 = 0
\]
Substituting the coefficients of the prolongation of the operator $Q$ into the above equation, we obtain the system of determining equation
\begin{gather*}
\tau_x = 0, \quad \tau_y = 0, \quad \tau_u = 0,\quad \tau_t = 3 \xi_x, \quad \tau_{tt}=0;  \\
\xi_y = 0, \quad \xi_u = 0, \quad \xi_{xx} = 0, \quad \xi_{xt} = 0; \\
\zeta_t = 0, \quad \zeta_x = 0, \quad \zeta_u = 0, \quad \zeta_y = \xi_x;\\
\eta_{xu} =0, \quad \eta_{yu} =0, \quad \eta_{uu} =0, \quad 2a\eta u + a(\tau_t-\xi_x)u^2 -\xi_t  = 0,\\
\eta_t  + a u^2 \eta_x + \eta_{xxx} + \eta_{xyy} = 0.
\end{gather*}
Solving the above system, we obtain
\[
\tau = c_1+c_2t,\quad \xi = \frac{1}{3}c_2x+c_3,\quad \zeta = \frac{1}{3}c_2y+c_4,\quad \eta = -\frac{1}{3}c_2u,
\]
where $c_1,\ldots,c_4$ are arbitrary constants.
In an analogous to the previous examples way we get the four-dimensional Lie algebra of infinitesimal symmetry generators
\[
\langle \p_t, \  \p_x, \  \p_y,\  3t\p_t+x\p_x+y\p_y-u\p_u \rangle.
\]
\end{example}

As in the above examples it appears that in the most of cases  computation of symmetries for a single equation
is an algorithmic and simple exercise,
which can be easily done by direct computing or by many standard mathematical software such as MAPLE, MATHEMATICA, MuLie and so on.
In contrast to this, for parametric classes of differential equations, an exhaustive investigation of symmetries is usually a very difficult task,
that requires to solve the so-called group classification problem.

\subsection{Group classification of classes of differential equations}

One of the most famous problems of group analysis of differential equations is the group classification
which is one of the symmetry methods used to choose physically relevant models
from parametric classes of systems of (partial or ordinary) differential equations.
The parameters can be constants or functions of independent variables, unknown functions and their derivatives.
Solving this problem is interesting not only from mathematical point of view, but is also important for applications.
In physical models there often exist a priori requirements to symmetry groups that follow from physical laws
(in particular, from the Galilei or relativistic theories).
Moreover, modeling differential equations could contain parameters or functions that have been found experimentally and so are not strictly fixed.
At the same time mathematical models have to be simple enough to analyze and solve them.
Solving the problems of group classification makes possible to accept for the criterion of applicability the following statement.
The modeling differential equation has to admit a group with certain properties or the richest symmetry group among the possible ones.

Exhaustive consideration of the problem of group classification for a parametric class $\mathcal{L}$
of systems of differential equations includes the following steps:
\begin{enumerate}
\item
Finding the group $G^{\cap}$
of local transformations that are symmetries for all systems from $\mathcal{L}$.
\item
Construction of the group $G^{\sim}$ (the equivalence group) of local transformations
which transform $\mathcal{L}$ into itself.
\item
Description of all possible $G^{\sim}$-inequivalent values of parameters that admit
maximal invariance groups wider than $G^{\cap}$.
\end{enumerate}
Following S.~Lie, one usually considers infinitesimal transformations instead of finite ones.
In such a way the problem of group classification can be simplified to a problem for Lie algebras of vector fields (infinitesimal generators of symmetry groups).
Thus the group classification in a class of differential equations is reduced to integration of an overdetermined system
of partial  differential equations with respect to both coefficients of infinitesimal symmetry operators and arbitrary elements.
That is why it is much more complicated problem than finding the Lie symmetry group of a single differential equation.

Below we present the classical algorithm of group classification restricting ourselves,
for simplicity, to the case of one differential equation of the form
\begin{equation}
\label{exameq}
L^\theta(x,u_{(n)})=L(x,u_{(n)},\theta(x,u_{(n)}))=0.
\end{equation}
Here
$x=(x_1,\ldots,x_l)$ denotes independent variables,
$u$ is a dependent variable,
$u_{(n)}$ is the set of all the partial derivatives of the function $u$ with respect to $x$
of order no greater than $n$, including $u$ as the derivative of zero order.
$L$ is a fixed function of $x$, $u_{(n)}$ and $\theta$.
$\theta$ denotes the set of of arbitrary (parametric) functions
$\theta(x,u_{(p)})=(\theta^1(x,u_{(p)}),\ldots,\theta^k(x,u_{(p)}))$
satisfying the conditions
\begin{equation}\label{examco}
S(x,u_{(p)},\theta_{(q)}(x,u_{(p)}))=0, \quad S=(S_1,\ldots,S_r).
\end{equation}
These conditions consist of $r$ differential equations on $\theta$,
where $x$ and $u_{(p)}$ play the role of independent variables
and $\theta_{(q)}$ stands for the set of all the partial derivatives of $\theta$ of order no greater than $q$.
In what follows we call the functions $\theta(x,u_{(p)})$ arbitrary elements.
Sometimes this set is additionally constrained
by the non-vanish condition $S'(x, u(p), \theta_{(q)}(x, u(p)))\ne 0$ with another
tuple $S'$ of differential functions.
Denote the class of equations of form~(\ref{exameq}) with the arbitrary elements $\theta$
satisfying the constraint~(\ref{examco}) as~$L|_S.$

\begin{example}
For the class~\eqref{eq2DimDispZKEq} of the generalized Zakharov--Kuznetsov equation,
in the above notation
$p=0$, the set of arbitrary elements $\theta(x,u_{(p)})$ consists of functions
$F$, $G$ and $H$  satisfying system of differential equations
\begin{equation}\label{sysEqsArbElem}
F_t=0, \quad F_x=0,\quad G_t=0,\quad G_x=0,\quad H_t=0, \quad H_x=0,
\end{equation}
and system of inequalities
\begin{equation}\label{sysIneqsArbElem}
F_u\ne0,\quad G_u\ne0, \quad H_{uu}\ne0.
\end{equation}
\end{example}

Let the functions $\theta$ be fixed.
Each one-parametric group of local point transformations
that leaves equation~(\ref{exameq}) invariant corresponds to an
infinitesimal symmetry operator of form
\[
Q=\xi^a(x,u)\p_{x_a}+\eta(x,u)\p_u.
\]
The complete set of such groups generates the principal group
$G^{\max}=G^{\max}(L,\theta)$ of equation~(\ref{exameq}).
The principal group $G^{\max}$ has a corresponding Lie algebra
$A^{\max}=A^{\max}(L,\theta)$ of infinitesimal  symmetry operators
of equation~(\ref{exameq}).
The group $G^{\cap}$
of local transformations that are symmetries for all systems is
\[
G^{\cap}=G^{\cap}(L,S)=\bigcap_{\theta:S(\theta)=0}G^{\max}(L,\theta)
\]
with the corresponding Lie algebra of form
\[
A^{\cap}=A^{\cap}(L,S)=\bigcap_{\theta:S(\theta)=0}A^{\max}(L,\theta).
\]
Let $G^{\sim}=G^{\sim}(L,S)$ denote the point transformations group preserving the
form of equations from~$L|_S$ (group of equivalence transformations).
In other words,  $G^{\sim}$ maps any equation from class~$L|_S$ to an equation (possibly, another one) from the same class.

\begin{note}
Sometimes one considers a subgroup instead of the complete equivalence group, e.g.,
subgroup of continuous transformations of the complete equivalence group.
This is because the group of continuous transformations can be easily found using infinitesimal method,
and to find the complete equivalence group one needs to apply more cumbersome and sophisticated direct method
of construction of equivalence transformations.
\end{note}

\begin{example}\label{ExampleOnEquivTransf}
To find the connected component of the
unity of the equivalence group~$G^{\sim}$ of class~\eqref{eq2DimDispZKEq}, i.e., the subgroup of continuous equivalence transformations,
we have to investigate Lie symmetries of the
system that consists of equation~\eqref{eq2DimDispZKEq} and
additional conditions~\eqref{sysEqsArbElem} subject to~\eqref{sysIneqsArbElem}.
In other words we must seek for an operator from $G^{\sim}$ in the form
\begin{equation}\label{OperatorEquivTr}
X=\tau \partial_t+\xi \partial_x+\zeta \partial_y+\eta
\partial_u+\pi \partial_F+\rho \partial_G+\theta \partial_H
\end{equation}
using the infinitesimal invariance criterion applied to the
system
\begin{equation}\label{sysForEquivTr}
\begin{array}{l}
u_t + (F(u))_{xxx} + (G(u))_{xyy} + (H(u))_x =0,\\[1ex]
F_t=F_x=F_y=0,\quad G_t=G_x=G_y=0,\quad H_t=H_x=H_y=0,
\end{array}
\end{equation}
subject to conditions~\eqref{sysIneqsArbElem}.
Here $u$, $F$, $G$ and $H$ are considered as differential variables:
$u$ on the space $(t,x,y)$ and $F$, $G$, $H$ on the extended space
$(t, x, y, u)$. The coordinates $\tau$, $\xi$, $\zeta$ and $\eta$ of
the operator~\eqref{OperatorEquivTr} are sought as functions of $t$,
$x$, $y$ and $u$ while the coordinates $\pi$, $\rho$ and $\theta$
are sought as functions of $t$, $x$, $y$, $u$, $F$, $G$, $H$.

\end{example}

The problem of group classification consists in finding all possible
inequivalent cases of extensions of $A^{\max}$, i.e.
in a listing all $G^{\sim}$-inequivalent  values of $\theta$ that satisfy
equation~(\ref{examco}) and the condition
$A^{\max}(L,\theta)\ne A^{\cap}$.

In the approach used here group classification is application
of the following algorithm~\mbox{\cite{Ovsiannikov1982}:}
\begin{enumerate}
\item
From the infinitesimal Lie invariance criterion~\cite{Ovsiannikov1982,Olver1986} we find the system of
determining equations for the coefficients of the infinitesimal generator~$Q$.
It is possible that some of the determining equations does not contain arbitrary elements
and therefore can be integrated immediately.
The equations containing arbitrary elements explicitly
are called classifying equations. %The main difficulty of group classification
%is the need to solve classifying equations with
%respect to the coefficients of the operator $Q$ and arbitrary
%elements simultaneously.
\item
We decompose the determining equations with respect
to all unconstrained derivatives of arbitrary elements.
This gives a system
of partial differential equations for coefficients of the
infinitesimal operator $Q$ only. Solving this system yields the algebra $A^{\cap}$ of point transformations
that are symmetries for all equations from~$L|_S.$
\item
To construct the equivalence group $G^{\sim}$ of the class~$L|_S$
one has to investigate the point symmetry transformations
of system~{(\ref{exameq}), (\ref{examco})}, considering it as a system
of partial differential equations with respect to $\theta$ with the
independent variables $x,\ u_{(n)}$.
If we restrict ourselves to studying the connected component of unity in $G^{\sim},$
(i.e., finding continuous equivalence transformations only),
the Lie infinitesimal method of finding symmetries of this system can be applied.
To find the complete  equivalence group
(including discrete transformations) one has to use the direct method.
\item
If $A^{\max}$ is an extension of $A^{\cap}$
(i.e. in the case $A^{\max}(L,\theta)\ne A^{\cap}$), then
the classifying equations define a system of nontrivial equations for
arbitrary elements~$\theta$.
Depending on their form and number, we obtain different cases of extensions of~$A^{\cap}$.
\end{enumerate}

\begin{note}
To integrate completely the determining equations
often it is necessary to investigate a large number of different cases of extensions of~$A^{\cap}$.
There exist different methods allowing to avoid cumbersome enumeration of possibilities in solving the determining
equations.
To solve determining equations for the coefficients of symmetry generators of Zakharov--Kuznetsov equations
we use a method which involves the investigation of compatibility of the classifying
equations~\cite{Nikitin&Popovych2001,Popovych&Ivanova2004NVCDCEs}.
\end{note}

The result of application of the above algorithm  is a list of equations with their
Lie invariance algebras.
The problem of group classification is assumed to be completely solved if
\begin{enumerate}%[\it i\rm)]
\item
the list contains all possible inequivalent cases of extensions;
\item\vspace{-1ex}
all equations from the list are mutually inequivalent with respect to
the transformations from $G^{\sim}$;
\item\vspace{-1ex}
the obtained algebras are the maximal invariance algebras of the respective equations.
\end{enumerate}
Such list may include equations being mutually equivalent with respect to
point transformations which do not belong to $G^{\sim}.$
Knowing such additional equivalences allows to simplify essentially
further investigation of~$L|_S$.
Constructing them sometimes is considered as the fifth step of the algorithm of group classification.
Then, the above enumeration of requirements to the resulting list of classification can be completed
by the following step:
\begin{enumerate}%[\it i\rm)]\setcounter{enumi}{3}
\setcounter{enumi}{3}
\item
all possible additional equivalences between the listed equations are constructed in
explicit form.
\end{enumerate}
See, e.g.,~\cite{Popovych&Ivanova2004NVCDCEs} for more details.

When the symmetry group is known, a wide range of applications becomes available, e.g.,
construction of invariant solutions of nonlinear equations.
Indeed, group analysis is one of very few systematic methods known for deducing exact solutions of nonlinear
partial differential equations.

\subsection{Invariant solutions of differential equations}

Although Lie symmetry analysis does not help to construct general solutions of systems of nonlinear PDEs
it often gives an approach to deduct wide classes of solutions being invariant with respect to different subgroups of the Lie symmetry group.
Roughly speaking, the main theorem on invariant solutions~\cite{Ovsiannikov1982,Olver1986} claims
that all solutions invariant with respect to $r$-parametric group of symmetries
(with some restrictions on the form of the algebra) of the given $n$-dimensional system
can be obtained by solving a system of differential equation with $n-r$ independent variables.
In particular, if $r=n-1$, invariant solutions can be constructed via solving a system of ordinary differential equations.

\begin{example}
Generalized Zakharov--Kuznetsov equations~\eqref{eq2DimDispZKEq} are three-dimensional.
Therefore, to reduce an equation of form~\eqref{eq2DimDispZKEq}
to ordinary differential equation we have to use two-dimensional subalgebras of its symmetry algebra.
Reduction of~\eqref{eq2DimDispZKEq} with respect to one-dimensional symmetry algebras provides us with the two-dimensional reduced equations.
Note, that the reduction with respect to the three-dimensional subalgebras leads to algebraic equations.
\end{example}

To construct solutions of a system of partial differential equations invariant with respect to $r$-dimensional symmetry algebra
spanned by symmetry generators $v_i=\xi^i_j\p_i+\eta_j\p_{u^j}$, $j=1,\ldots,r$, we need to solve a system
of $r$ first-order PDEs:
\[
\xi^i_ju_{i}=\eta_j, \quad j=1,\ldots, n .
\]
Solution of this system provides us with expressions for $n-r$ new independent variables and Ansatz for the dependent variables.
Substituting this to the initial system we obtain a system of differential equations with $n-r$ independent variables.
A detailed example of implementation of this method will be given in section~\ref{SectionExactSol}.

\begin{note}
The above mentioned procedure works only if the symmetry algebra satisfies the property of transversality.
For more details see~\cite{Ovsiannikov1982,Olver1986}.
\end{note}

As we have already noticed, in general ``almost every'' subgroup of a Lie symmetry group corresponds to a class of invariant solutions.
Since almost always there exist an infinite number of such subgroups, often it is practically impossible to list all invariant solutions.
Therefore one needs an effective systematic tool of their classification
that gives an ``optimal system'' of such solutions, from where one can find all possible invariant solutions.

Any two conjugate subgroups of a Lie symmetry group of a system of PDEs give rise to reduced equations
that are related by a conjugacy transformation
in the point symmetry group of the system
acting on the invariant solutions determined by each subgroup.
Hence, up to the action of the point symmetry transformations,
all invariant solutions for a given system
can be obtained by selecting a subgroup
in each conjugacy class of all admitted point symmetry subgroups.
Such a selection is called an optimal set of subgroups~\cite{Ovsiannikov1982}.
A set of subalgebras of the Lie symmetry algebra corresponding to the optimal set of subgroups consists of
subalgebras inequivalent with respect to the action of adjoint representation of the Lie symmetry group on its Lie
algebra.

An effective algorithm of construction of optimal systems of subalgebras of Lie algebras is given in~\cite{Ovsiannikov1982}
(see also a simpler explanation and examples for one-dimensional subalgebras in~\cite{Olver1986}).
In section~\ref{SectionExactSol} we illustrate this algorithm by examples of the usual and modified Zakharov--Kuznetsov equations.

\section{Equivalence transformations of the GZK equations}\label{SectionEquivTrasf}

We start the group classification procedure of class~\eqref{eq2DimDispZKEq} of the GZK equations
from investigating its group of equivalence transformations.

First, we describe construction of the continuous equivalence transformations.
As it is stated in Example~\ref{ExampleOnEquivTransf}, to reach this goal one has to investigate Lie symmetries of system~\eqref{sysForEquivTr}.
Application of the infinitesimal invariance criterion to system~\eqref{sysForEquivTr} yields the
following determining equations for $\tau$, $\xi$, $\zeta$, $\eta$,
$\pi$, $\rho$ and $\theta$:
\begin{gather}\nonumber
\tau_x=\tau_y=\tau_u=0,\quad \xi_y=\xi_u=\xi_{xx}=0,\\ \nonumber
\zeta_t=\zeta_x=\zeta_u=\zeta_{yy}=0, \quad \eta_{t}=\eta_{x}=\eta_{y}=0,\\
\nonumber \pi_t=\pi_x=\pi_y=\pi_u=\pi_G=\pi_H=\pi_{FF}=0,\quad
\pi_F-\eta_u+\tau_t-3\xi_x=0,\\
\nonumber
\rho_t=\rho_x=\rho_y=\rho_u=\rho_F=\rho_H=\rho_{GG}=0,\quad
\rho_G-\eta_u+\tau_t-\xi_x-2\zeta_y=0,\\
\label{sysDetEqEquivTr}
\theta_t=\theta_x=\theta_y=\theta_F=\theta_G=0,\quad
\theta_u-\xi_t=0\quad \theta_H-\eta_u+\tau_t-\xi_x=0.
\end{gather}
After easy calculations from~\eqref{sysDetEqEquivTr}, we find
the coefficients of the infinitesimal operators of continuous equivalent transformations of class \eqref{eq2DimDispZKEq} have the form
\[
\begin{array}{l}
\tau=c_1+c_5t,\quad \xi=c_2+c_6x+c_9t,\quad \zeta=c_3+c_7y,\quad
\eta=c_4+c_8u,\\
\pi=(c_5-3c_6-c_8)F+c_{11},\quad \rho=(c_5-c_6-2c_7-c_8)G+c_{12},\\
\theta=(c_5-c_6-c_8)H+c_9u+c_{10}.
\end{array}
\]
Then the Lie algebra of the equivalence group~$G^{\sim}$ for
class~\eqref{eq2DimDispZKEq} is
\[
\begin{array}{l}
A^{\sim}=\langle\partial_t,\,\partial_x,\, \partial_y,\,
\partial_u,\,  t\partial_t+F\partial_F+G\partial_G+H\partial_H,\,
x\partial_x-3F\partial_F-G\partial_G-H\partial_H,\\
y\partial_y-2G\partial_G,\,
u\partial_u-F\partial_F-G\partial_G-H\partial_H,\,
 t\partial_x+u\partial_H,\, \p_F,\,\p_G,\, \partial_H\rangle .
\end{array}
\]

Now, to recover the formulas for the transformations of variables one needs to solve the so-called Lie equations (see, e.g.,~\cite{Olver1986,Ovsiannikov1982}).
Namely, for any infinitesimal operator $X=\tau \partial_t+\xi \partial_x+\zeta \partial_y+\eta
\partial_u+\pi \partial_F+\rho \partial_G+\theta \partial_H$ the corresponding finite transformations of variables can be found from solving the system
\begin{gather*}
\frac{d\tilde t}{d\varepsilon}=\tau, \quad \tilde t|_{\varepsilon=0}=t,\\
\frac{d\tilde x}{d\varepsilon}=\xi, \quad \tilde x|_{\varepsilon=0}=x,\\
\ldots\\
\frac{d\tilde H}{d\varepsilon}=\theta, \quad \tilde H|_{\varepsilon=0}=H.
\end{gather*}
Solving this system for every infinitesimal equivalence generator and gathering all 12 results together,
we get that the group~$G^{\sim}_{\rm cont}$ of \emph{continuous} equivalence transformations of class~\eqref{eq2DimDispZKEq} is as follows:
\begin{gather*}
\tilde t=e^{\varepsilon_1}t+\varepsilon_6,\quad
\tilde x=e^{\varepsilon_2}x+e^{\varepsilon_5}t+\varepsilon_7,\quad
\tilde y=e^{\varepsilon_3}y+\varepsilon_8,\quad
\tilde u=e^{\varepsilon_4}u+\varepsilon_9,\\
\tilde F=e{-\varepsilon_1+3\varepsilon_2+\varepsilon_4}F+\varepsilon_{10},\quad
\tilde G=e^{-\varepsilon_1+\varepsilon_2+2\varepsilon_3+\varepsilon_4}G+\varepsilon_{11},\\
\tilde H=e^{-\varepsilon_1+\varepsilon_2+\varepsilon_4}H+e^{-\varepsilon_1+\varepsilon_4+\varepsilon_5}u+\varepsilon_{12}.
\end{gather*}

To find the \emph{complete} equivalence group of class~\eqref{eq2DimDispZKEq} we use the direct method.
More precisely, we look for transformations of form
\[
\tilde t=\tilde t(t,x,y,u),\quad \tilde x=\tilde x(t,x,y,u), \quad \tilde y=\tilde y(t,x,y,u), \quad \tilde u=\tilde u(t,x,y,u)
\]
that relates equations $u_t+(F(u))_{xxx}+(G(u))_{xyy}+(H(u))_x=0$ and
$\tilde u_{\tilde t}+(\tilde F(\tilde u))_{\tilde x\tilde x\tilde x}+(\tilde G(\tilde u))_{\tilde x\tilde y\tilde y}+(\tilde H(\tilde u))_{\tilde x}=0$.

Now we have to express the non-transformed variables in terms of the ``tilded'' ones and substitute them into equation~\eqref{eq2DimDispZKEq}.
Requiring that the obtained equation belongs to class~\eqref{eq2DimDispZKEq}
we get a polynomial equation with respect to the derivatives of~$u$.
Setting to zero its coefficients with respect to the unconstrained variables
we obtain a system of overdetermined partial differential equations, general solution of which gives that the most general form of
the transformation from the complete equivalence group~$G^{\sim}$ of class~\eqref{eq2DimDispZKEq} is
\begin{gather*}
\tilde t=\varepsilon_1t+\varepsilon_6,\quad
\tilde x=\varepsilon_2x+\varepsilon_5t+\varepsilon_7,\quad
\tilde y=\varepsilon_3y+\varepsilon_8,\quad
\tilde u=\varepsilon_4u+\varepsilon_9,\\
\tilde F=\varepsilon_1^{-1}\varepsilon_2^3\varepsilon_4F+\varepsilon_{10},\quad
\tilde G=\varepsilon_1^{-1}\varepsilon_2\varepsilon_3^2\varepsilon_4G+\varepsilon_{11},\quad
\tilde H=\varepsilon_1^{-1}\varepsilon_2\varepsilon_4H+\varepsilon_1^{-1}\varepsilon_4\varepsilon_5u+\varepsilon_{12},
\end{gather*}
where $\varepsilon_1,\ldots,\varepsilon_{12}$ are arbitrary constants, $\varepsilon_1\varepsilon_2\varepsilon_3\varepsilon_4\ne0$.

For more details and examples of application of the direct method of finding of equivalence transformations see, e.g.,~\cite{Kingston&Sophocleous1998}.

\begin{note}
As one can see, the only discrete equivalence transformations for class~\eqref{eq2DimDispZKEq} are alternating of signs of the
dependent and independent variables (extended to the arbitrary elements).
However, for many classes this is not the case, i.e., there can exist nontrivial complicated discrete equivalence transformations
that can essentially simplify classification problem (see, e.g.,~\cite{Huang&Ivanova2007,Huang&Zhou2011,Ivanova&Popovych&Sophocleous2010} and references therein).
\end{note}

\section{Lie symmetries of the GZK equations}\label{SectionLieSymGZKEq}

We search for infinitesimal generators of Lie symmetries of equations of class~\eqref{eq2DimDispZKEq} in form
\[
Q=\tau(t,x,y,u)\p_t+\xi(t,x,y,u)\p_x+\zeta(t,x,y,u)\p_y+\eta(t,x,y,u)\p_u.
\]
From the infinitesimal invariance criterion we obtain the following system of the determining equations for the coefficients of the infinitesimal generators:
\begin{gather}
\tau_x=\tau_y=\tau_u=\xi_y=\xi_u=\zeta_t=\zeta_x=\zeta_u=\eta_{uu}=0, \nonumber\\
(\eta_xG_u)_u=0, \nonumber\\
\eta_t+F_u\eta_{xxx}+G_u\eta_{xyy}+\eta_xH_u = 0, \nonumber\\
(F_u\eta_x)_u-F_u\xi_{xx}=0, \nonumber\\
2(\eta_yG_u)_u-G_u\zeta_{yy}=0, \nonumber\\
\eta F_{uu}+(\tau_t-3\xi_x)F_u = 0, \nonumber\\
\eta G_{uu}+(\tau_t-\xi_x-2\zeta_y)G_u = 0, \nonumber\\
3(\eta_{xx}F_u)_u+(\eta_{yy}G_u)_u+\eta H_{uu}-\xi_{xxx}F_u+(\tau_t-\xi_x)H_u-\xi_t= 0. \label{sysDetEqZKEq}
\end{gather}

Solving this system up to the equivalence group~$G^{\sim}$ we obtain the complete group classification of class~\eqref{eq2DimDispZKEq}
that presented in Table~\ref{TableGrCl2DimDispZKEq}.

\newpage

\begin{center}
\footnotesize\setcounter{tbn}{0}\renewcommand{\arraystretch}{1.1}
Table \refstepcounter{tabu}\label{TableGrCl2DimDispZKEq}\thetabu. Group classification of class~\eqref{eq2DimDispZKEq}. \\[1ex]
\begin{tabular}{|l|c|c|c|l|}
\hline
N\vspacebefore & $F(u)$ & $G(u)$ & $H(u)$ & \hfill  $A^{\rm max}\hfill$\\[0.5ex]
\hline
\refstepcounter{tbn}\label{caseFarbitraryGarbitraryHarbitrary}\thetbn.\vspacebefore &  $\forall$ &  $\forall$ & $\forall$ &
    $A^{\cap}=\langle \p_t,\, \p_x,\, \p_y \rangle $ \\[0.5ex]
%1
\refstepcounter{tbn}\label{caseFukGumHun}\thetbn.\vspacebefore &  $u^k$ & $gu^m$ &  $hu^n$
            & $A^{\cap}+\langle (k+2-3n)t\p_t+(k-n)x\p_x+(m-n)y\p_y+2u\p_u \rangle $ \\[0.5ex]
\refstepcounter{tbn}\label{caseFukGumHulnu}\thetbn.\vspacebefore &  $u^k$ & $gu^m$ &  $hu\ln u$ &
           $A^{\cap}+\langle (k-1)t\p_t+(2ht+kx-x)\p_x+(m-1)y\p_y+2u\p_u \rangle $ \\[0.5ex]
\refstepcounter{tbn}\label{caseFukGumHlnu}\thetbn.\vspacebefore &  $u^k$ & $gu^m$ &  $h\ln u$ &
           $A^{\cap}+\langle (k+2)t\p_t+kx\p_x+my\p_y+2u\p_u \rangle $ \\[0.5ex]
\refstepcounter{tbn}\label{caseFlnuGumHun}\thetbn.\vspacebefore &  $\ln u$ & $gu^m$ &  $hu^n$
          & $A^{\cap}+\langle (3n-2)t\p_t+nx\p_x+(n-m)y\p_y-2u\p_u \rangle $ \\[0.5ex]
\refstepcounter{tbn}\label{caseFlnuGumHulnu}\thetbn.\vspacebefore &  $\ln u$ & $gu^m$ &  $hu\ln u$
          & $A^{\cap}+\langle t\p_t+(x-2ht)\p_x-(m-1)y\p_y-2u\p_u \rangle $ \\[0.5ex]
\refstepcounter{tbn}\label{caseFlnuGumHlnu}\thetbn.\vspacebefore &  $\ln u$ & $gu^m$ &  $h\ln u$
          & $A^{\cap}+\langle 2t\p_t+my\p_y+2u\p_u \rangle $ \\[0.5ex]
\refstepcounter{tbn}\label{caseFekuGemuHenu}\thetbn.\vspacebefore &  $e^{ku}$ & $ge^{mu}$ &  $he^{nu}$
         & $A^{\cap}+\langle (k-3n)t\p_t+(k-n)x\p_x+(m-n)y\p_y+2\p_u \rangle $ \\[0.5ex]
\refstepcounter{tbn}\label{caseFekuGemuHu2}\thetbn.\vspacebefore &  $e^{ku}$ & $ge^{mu}$ &  $hu^2$
         & $A^{\cap}+\langle kt\p_t+(kx+4ht)\p_x+my\p_y+2\p_u \rangle $ \\[0.5ex]
\refstepcounter{tbn}\label{caseFuGemuHenu}\thetbn.\vspacebefore &  $u$ & $ge^{mu}$ &  $he^{nu}$
         & $A^{\cap}+\langle 3nt\p_t+nx\p_x+(n-m)y\p_y-2\p_u \rangle $ \\[0.5ex]
\refstepcounter{tbn}\label{caseFuGemuHu2}\thetbn.\vspacebefore &  $u$ & $ge^{mu}$ &  $hu^2$
         & $A^{\cap}+\langle 4ht\p_x+my\p_y+2\p_u \rangle $ \\[0.5ex]
\refstepcounter{tbn}\label{caseFukGlnuHun}\thetbn.\vspacebefore &  $u^k$ & $g\ln u$ &  $hu^n$
         & $A^{\cap}+\langle (k-3n+2)t\p_t+(k-n)x\p_x-ny\p_y+2u\p_u \rangle $ \\[0.5ex]
\refstepcounter{tbn}\label{caseFukGlnuHlnu}\thetbn.\vspacebefore &  $u^k$ & $g\ln u$ &  $h\ln u$
         & $A^{\cap}+\langle (k+2)t\p_t+kx\p_x+2u\p_u \rangle $ \\[0.5ex]
\refstepcounter{tbn}\label{caseFukGlnuHulnu}\thetbn.\vspacebefore &  $u^k$ & $g\ln u$ &  $hu\ln u$
         & $A^{\cap}+\langle (k-1)t\p_t+(2ht+kx-x)\p_x-y\p_y+2u\p_u \rangle $ \\[0.5ex]
\refstepcounter{tbn}\label{caseFlnuGlnuHun}\thetbn.\vspacebefore &  $\ln u$ & $g\ln u$ &  $hu^n$
         & $A^{\cap}+\langle (3n-2)t\p_t+nx\p_x+ny\p_y-2u\p_u \rangle $ \\[0.5ex]
\refstepcounter{tbn}\label{caseFlnuGlnuHlnu}\thetbn.\vspacebefore &  $\ln u$ & $g\ln u$ &  $h\ln u$
         & $A^{\cap}+\langle t\p_t+u\p_u \rangle $ \\[0.5ex]
\refstepcounter{tbn}\label{caseFlnuGlnuHulnu}\thetbn.\vspacebefore & $\ln u$ & $g\ln u$ &  $hu\ln u$
         & $A^{\cap}+\langle t\p_t+(x-2ht)\p_x+y\p_y-2u\p_u \rangle $ \\[0.5ex]
\refstepcounter{tbn}\label{caseFekuGuHenu}\thetbn.\vspacebefore & $e^{ku}$ & $gu$ &  $he^{nu}$
         & $A^{\cap}+\langle (k-3n)t\p_t+(k-n)x\p_x-ny\p_y+2\p_u \rangle $ \\[0.5ex]
\refstepcounter{tbn}\label{caseFekuGuHu2}\thetbn.\vspacebefore & $e^{ku}$ & $gu$ &  $hu^2$
         & $A^{\cap}+\langle kt\p_t+(kx+4ht)x\p_x+2\p_u \rangle $ \\[0.5ex]
\refstepcounter{tbn}\label{caseFuGuHu2}\thetbn.\vspacebefore &  $u$ & $gu$ &  $hu^2$
         & $A^{\cap}+\langle 3t\p_t+x\p_x+y\p_y-2u\p_u,\, 2ht\p_x+\p_u \rangle $ \\[0.5ex]
\refstepcounter{tbn}\label{caseFuGu-13Hu-13}\thetbn.\vspacebefore &  $u$ & $u^{-1/3}$ &  $u^{-1/3}$
         & $A^{\cap}+\langle 6t\p_t+2x\p_x+3u\p_u,\, \cos{2y}\p_y+3u\sin{2y}\p_u,$ \\ &&&&
          $\sin{2y}\p_y-3u\cos{2y}\p_u \rangle $ \\[0.5ex]
\refstepcounter{tbn}\label{caseFuG-u-13H-u-13}\thetbn.\vspacebefore &  $u$ & $-u^{-1/3}$ &  $-u^{-1/3}$
         & $A^{\cap}+\langle 6t\p_t+2x\p_x+3u\p_u,\, \cos{2y}\p_y+3u\sin{2y}\p_u,$ \\ &&&&
          $\sin{2y}\p_y-3u\cos{2y}\p_u \rangle $ \\[0.5ex]
\refstepcounter{tbn}\label{caseFuG-u-13Hu-13}\thetbn.\vspacebefore &  $u$ & $-u^{-1/3}$ &  $u^{-1/3}$
           & $A^{\cap}+\langle 6t\p_t+2x\p_x+3u\p_u,\, e^{2y}(\p_y-3u\p_u),\, e^{-2y}(\p_y+3u\p_u) \rangle $ \\[0.5ex]
\refstepcounter{tbn}\label{caseFuGu-13H-u-13}\thetbn.\vspacebefore &  $u$ & $u^{-1/3}$ &  $-u^{-1/3}$
           & $A^{\cap}+\langle 6t\p_t+2x\p_x+3u\p_u,\, e^{2y}(\p_y-3u\p_u),\, e^{-2y}(\p_y+3u\p_u) \rangle $ \\[0.5ex]
\hline
\end{tabular}
\end{center}
{\footnotesize
Here $g=\pm1$, $h=\pm1$.
}

\begin{proof}
To obtain the classification result we need to solve the system of determining equations~\eqref{sysDetEqZKEq}.
Integrating the equations that do not contain arbitrary elements we get
\begin{equation}\label{CoefsSymIntegr}
\tau =\tau(t),\quad \xi=\xi(t, x),\quad \zeta= \zeta(y), \quad \eta=\eta^1(t, x, y)u+\eta^0(t, x, y).
\end{equation}

Splitting the rest of the system~\eqref{sysDetEqZKEq}
with respect to the arbitrary elements and their non-vanishing
derivatives gives the equations $\tau_t=0$, $\xi_t=\xi_x=0$, $\zeta_y= 0$ and $\eta=0$ for
the coefficients of the operators from~$A^{\cap}$
of~\eqref{eq2DimDispZKEq}. As a result, we get case 1 of Table~1.

As one can see,
for any symmetry operator, equations
$\eta F_{uu}+(\tau_t-3\xi_x)F_u = 0$ and $\eta G_{uu}+(\tau_t-\xi_x-2\zeta_y)G_u = 0$
give some equations (not greater than 3 each) on $F$ and $G$ of the general form
\[
(au+b)F_{uu}+cF_u=0,\quad (au+b)G_{uu}+fG_u=0,
\]
where $a$, $b$, $c$ and $f$ are constants. Solving this
system up to $G^{\sim}$, we obtain the following different values of $F(u)$ and $G(u)$:
$F(u)=u^k$, $F(u)=e^{ku}$ or $F(u)=\ln u \mod G^{\sim}$ and $G(u)=g u^m$, or $G(u)=ge^{mu}$ or $G(u)=g\ln u \mod G^{\sim}$.
Therefore, to complete the classification we have to consider
all possible combinations of the values of $F(u)$ and $G(u)$.
We attempted to present our calculations in reasonable
detail so that verification would be feasible. %Below we set $\eta(t,x,y,u)=\eta^1(t,x,y)u+\eta^0(t,x,y)$.

(I) $F=u^{k}$ and $G=gu^m$ mod $G^{\sim}$, where $m \neq 0$, $g=\pm 1$.
Substituting $F$, $G$ and~\eqref{CoefsSymIntegr} into the rest of the equations~\eqref{sysDetEqZKEq} yields
\begin{equation}\label{sysDetEqFukGgum}
\begin{array}{ll}
\eta^1_x=\xi_{xx}=0, \quad (m-1)\eta^0_x=(m-1)\eta^0_y=(k-1)\eta^0_x=0, \quad (m-1)\eta^0=(k-1)\eta^0=0,\\
2m\eta^1_y-\zeta_{yy}=0,\quad (k-1)\eta^1+(\tau_t-3\xi_x)=0,\quad (m-1)\eta^1+(\tau_t-\xi_x-2\zeta_y)=0.
\end{array}
\end{equation}
From the last three equations of this system we can get two classifying conditions:
\begin{equation}\label{ClassiConFukGgum}
(k-1)\eta^1_y=0,\quad (3m+1)\eta^1_y=0,
\end{equation}
which can be decomposed
into four cases:
\[
(i)\ k\neq1, m\neq -\frac{1}{3}; \quad(ii)\ k\neq1, m= -\frac{1}{3}; \quad (iii)\ k=1, m\neq -\frac{1}{3};\quad (iv)\ k=1, m= -\frac{1}{3}.
\]

(i) For this case equations~\eqref{sysDetEqFukGgum} and \eqref{ClassiConFukGgum} imply $\eta^1_x=\eta^1_y=\xi_{xx}=\eta^0=0$.
Hence, the equation for $H$ in~\eqref{sysDetEqZKEq} looks like $\mu uH_{uu}+\nu H_u+\omega=0$, where $\mu, \nu, \omega=\const$.
Therefore, $H(u)$ must take one of the following three values: $H(u)=hu^n$ or $H(u)=hu\ln u$ or $H(u)=h\ln u \mod G^{\sim}$, where $h=\pm 1$.
Substituting these three values in~\eqref{sysDetEqZKEq} and solving the obtained equations,
we get  cases \ref{caseFukGumHun}, \ref{caseFukGumHulnu} and \ref{caseFukGumHlnu} respectively.

(ii) From $k\neq 1$ and \eqref{sysDetEqFukGgum} and \eqref{ClassiConFukGgum} we obtain $\eta^1_x=\eta^1_y=\xi_{xx}=\eta^0=0$
that leads to the subcases of the previously obtained cases \ref{caseFukGumHun}, \ref{caseFukGumHulnu} and \ref{caseFukGumHlnu}.

(iii) With the restriction $m\neq 1$, we have again $\eta^1_x=\eta^1_y=\xi_{xx}=\eta^0=0$.
This also leads to the subcases of cases \ref{caseFukGumHun}, \ref{caseFukGumHulnu} and \ref{caseFukGumHlnu}.

If $m=1$, then from \eqref{sysDetEqFukGgum} and \eqref{ClassiConFukGgum} we have  $\eta^1_x=\eta^1_y=\xi_{xx}=0$.
Thus, from the equation for $H(u)$ we deduce that it takes
one of the following three values: $H(u)=hu^n$ or $H(u)=hu\ln u$ or $H(u)=h\ln u \mod G^{\sim}$, where $h=\pm 1$.
Solving the rest of equations~\eqref{sysDetEqZKEq}  and \eqref{sysDetEqFukGgum} with $H(u)=hu^n$,
we can find that the only extension case corresponds to the values $n=2$ and is tabulated as case~\ref{caseFuGuHu2}.
The values $H(u)=hu\ln u$ and $H(u)=h\ln u$ give the classification cases  \ref{caseFukGumHulnu} and \ref{caseFukGumHlnu} with $k=m=1$ respectively.

(iv) After obvious simple computations, we get cases \ref{caseFuGu-13Hu-13}--\ref{caseFuGu-13H-u-13}.

(II) $F=u^{k}$ and $G=ge^{mu}\!\!\mod\! G^{\sim}$, where $m \neq 0$, $g=\pm 1$. Substituting $F$, $G$ and~\eqref{CoefsSymIntegr}
into~\eqref{sysDetEqZKEq} we derive
\begin{equation}\label{sysDetEqFukGgemu}
\begin{array}{ll}
\eta^1=\xi_{xx}=0, \quad \eta^0_x=\eta^0_t=\eta^0_{yy}=0, \quad k(k-1)\eta^0=0,\\
m\eta^0_y-\zeta_{yy}=0,\quad \tau_t-3\xi_x=0,\quad m\eta^0+(\tau_t-\xi_x-2\zeta_y)=0,\\
\eta^0H_{uu}+(\tau_t-\xi_x)H_u-\xi_t=0.
\end{array}
\end{equation}
If $k\neq1$, then $\eta^0=0$. Therefore, $H=\const$ that
contradicts with the assumption $H_u \neq 0$.
If $k=1$, then the last equation of system
\eqref{sysDetEqFukGgemu} looks like $\mu H_{uu}+\nu H_u+\omega=0$ with
respect to $H$, where $\mu, \nu, \omega=\const$. Thus, $H(u)$ must take
one of the following two values: $H(u)=he^{nu}$ or $H(u)=hu^2 \mod G^{\sim}$, where $h=\pm 1$.
Substituting these two values into system \eqref{sysDetEqFukGgemu} and solving the obtained equations
we get cases \ref{caseFuGemuHenu} and \ref{caseFuGemuHu2} respectively.

(III) $F=u^{k}$ and $G=g\ln u$ mod $G^{\sim}$, where $g=\pm 1$.
System~\eqref{sysDetEqZKEq} takes now the form
\begin{equation}\label{sysDetEqFukGglnu}
\begin{array}{ll}
\eta^1_t=\eta^1_x=\eta^1_y=\xi_{xx}=0, \quad \eta^0=0, \quad  \zeta_{yy}=0,\\
(k-1)\eta^1+\tau_t-3\xi_x=0,\quad -\eta^1+(\tau_t-\xi_x-2\zeta_y)=0,\\
\eta^1uH_{uu}+(\tau_t-\xi_x)H_u-\xi_t=0.
\end{array}
\end{equation}
The last equation of system \eqref{sysDetEqFukGglnu} looks like $\mu u H_{uu}+\nu H_u+\omega=0$ with
respect to $H$, where $\mu, \nu, \omega=\const$.
Thus, $H(u)$ must be of one of the following three values: $H(u)=hu^n$ or $H(u)=h\ln u$ or $H(u)=hu\ln u$ $\mod G^{\sim}$, where $h=\pm 1$.
Solving system \eqref{sysDetEqFukGglnu} fot these three values of $H$,
we can obtain cases \ref{caseFukGlnuHun}, \ref{caseFukGlnuHlnu}  and \ref{caseFukGlnuHulnu} respectively.

(IV) $F=e^{ku}$ and $G=gu^m$ mod $G^{\sim}$, where $m \neq 0$, $g=\pm 1$. From~\eqref{sysDetEqZKEq} we derive
\begin{equation}\label{sysDetEqFekuGgum}
\begin{array}{ll}
\eta^1=0, \quad \xi_{xx}=\eta^0_t=\eta^0_x=\eta^0_y=0, \quad  \zeta_{yy}=0, \quad m(m-1)\eta^0=0,\\
k\eta^0+\tau_t-3\xi_x=0,\quad \tau_t-\xi_x-2\zeta_y=0,\\
\eta^0H_{uu}+(\tau_t-\xi_x)H_u-\xi_t=0.
\end{array}
\end{equation}
The fourth and the last equations of system
\eqref{sysDetEqFekuGgum} imply  $m=1$, or $H=\const$ mod $G^{\sim}$ which
contradicts with the assumption that $H_u \neq 0$.  Thus, $H(u)$ must take
one of the following two values: $H(u)=he^{nu}$ or $H(u)=hu^2$ $\mod G^{\sim}$, where $h=\pm 1$.
Substituting these values of $H$ into system \eqref{sysDetEqFekuGgum} and solving it,
we obtain cases \ref{caseFekuGuHenu} and \ref{caseFekuGuHu2} respectively.

(V) $F=e^{ku}$ and $G=ge^{mu}$ mod $G^{\sim}$, $m \neq 0$, $g=\pm 1$.
From~\eqref{sysDetEqZKEq} we derive
\begin{equation}\label{sysDetEqFekuGgemu}
\begin{array}{ll}
\eta^1=0, \quad \xi_{xx}=\eta^0_t=\eta^0_x=\eta^0_y=0, \quad  \zeta_{yy}=0,\\
k\eta^0+\tau_t-3\xi_x=0,\quad m\eta^0+\tau_t-\xi_x-2\zeta_y=0,\\
\eta^0H_{uu}+(\tau_t-\xi_x)H_u-\xi_t=0.
\end{array}
\end{equation}
The last equation of system
\eqref{sysDetEqFekuGgemu} looks like $\mu H_{uu}+\nu H_u+\omega=0$ with
respect to $H$, where $\mu, \nu, \omega=\const$.
Thus, up to $G^{\sim}$, $H(u)$ must take
one of the following two values: $H(u)=he^{nu}$ or $H(u)=hu^2$, where $h=\pm 1$.
Now from system \eqref{sysDetEqFekuGgemu} we obtain easily cases \ref{caseFekuGemuHenu} and \ref{caseFekuGemuHu2}.

(VI) $F=\ln u$ and $G=ge^{mu}$ mod $G^{\sim}$,  $m \neq 0$, $g=\pm 1$. From~\eqref{sysDetEqZKEq} we derive
\begin{equation}\label{sysDetEqFlnuGgum}
\begin{array}{ll}
\eta^0=0, \quad \xi_{xx}=\eta^1_t=\eta^1_x=\eta^1_y=0, \quad  \zeta_{yy}=0,\\
-\eta^1+\tau_t-3\xi_x=0,\quad (m-1)\eta^1+\tau_t-\xi_x-2\zeta_y=0,\\
\eta^1uH_{uu}+(\tau_t-\xi_x)H_u-\xi_t=0.
\end{array}
\end{equation}
The last equation of system
\eqref{sysDetEqFlnuGgum} looks like $\mu uH_{uu}+\nu H_u+\omega=0$ with
respect to $H$, where $\mu, \nu, \omega=\const$.  Thus, $H(u)$ must take
one of the following three values: $H(u)=hu^n$ or $H(u)=hu\ln u$ or $H(u)=h\ln u$ $\mod G^{\sim}$, where $h=\pm 1$.
Solving now system \eqref{sysDetEqFlnuGgum} we obtain cases \ref{caseFlnuGumHun}, \ref{caseFlnuGumHulnu}  and \ref{caseFlnuGumHlnu} respectively.

(VII) $F=\ln u$ and $G=g\ln u$ mod $G^{\sim}$, where $g=\pm 1$. Substituting $F$, $G$ and~\ref{CoefsSymIntegr} into~\eqref{sysDetEqZKEq} we derive
\begin{equation}\label{sysDetEqFlnuGglnu}
\begin{array}{ll}
\eta^0=0, \quad \xi_{xx}=\eta^1_t=\eta^1_x=\eta^1_y=0, \quad  \zeta_{yy}=0,\\
-\eta^1+\tau_t-3\xi_x=0,\quad -\eta^1+\tau_t-\xi_x-2\zeta_y=0,\\
\eta^1uH_{uu}+(\tau_t-\xi_x)H_u-\xi_t=0.
\end{array}
\end{equation}
Similar to (VI), from the last equation of system \eqref{sysDetEqFlnuGglnu} we obtain that $H(u)$ must take
one of the following three values: $H(u)=hu^n$ or $H(u)=h\ln u$ or $H(u)=hu\ln u$ mod $G^{\sim}$, where $h=\pm 1$.
These values correspond to cases \ref{caseFlnuGlnuHun}, \ref{caseFlnuGlnuHlnu}  and \ref{caseFlnuGlnuHulnu} respectively.

The problem of group classification of class~\eqref{eq2DimDispZKEq} is completely solved.
\end{proof}

\begin{note}
Although equations~\eqref{eq2DimDispZKEq} with $H_{uu}=0$  are of low physical interest,
below for completeness we adduce the result of their symmetry classification.
\end{note}

If $H_{uu}=0$, equation~\eqref{eq2DimDispZKEq} is equivalent to
\begin{equation}\label{eq2DimDispZKEqH0}
u_t+(F(u))_{xxx}+(G(u))_{xyy}=0.
\end{equation}
Result of group classification of class~\eqref{eq2DimDispZKEqH0}
(up to the restriction of the equivalence group~~$G^{\sim}$ of class~\eqref{eq2DimDispZKEq} to $H(u)=0$) is presented in Table~\ref{TableGrCl2DimDispZKEqH0}.

\begin{center}
\footnotesize\setcounter{tbn}{0}\renewcommand{\arraystretch}{1.1}
Table \refstepcounter{tabu}\label{TableGrCl2DimDispZKEqH0}\thetabu. Group classification of class~\eqref{eq2DimDispZKEqH0}. \\[1ex]
\begin{tabular}{|l|c|c|l|}
\hline
N\vspacebefore & $F(u)$ & $G(u)$ &  \hfill  $A^{\rm max}\hfill$\\[0.5ex]
\hline
\refstepcounter{tbn}\label{caseFarbitraryGarbitraryH0}\thetbn.\vspacebefore &  $\forall$ &  $\forall$ &
    $A^{\cap}_0=\langle \p_t,\, \p_x,\, \p_y,\, 3t\p_t+x\p_x+y\p_y \rangle $ \\[0.5ex]
%1
\refstepcounter{tbn}\label{caseFukGumH0}\thetbn.\vspacebefore &  $u^k$ & $gu^m$ &
           $A^{\cap}_0+\langle (k+2)t\p_t+kx\p_x+my\p_y+2u\p_u \rangle $ \\[0.5ex]
\refstepcounter{tbn}\label{caseFlnuGumH0}\thetbn.\vspacebefore &  $\ln u$ & $gu^m$
          & $A^{\cap}_0+\langle 2t\p_t+my\p_y+2u\p_u \rangle $ \\[0.5ex]
\refstepcounter{tbn}\label{caseFekuGemuH0}\thetbn.\vspacebefore &  $e^{ku}$ & $ge^{mu}$
         & $A^{\cap}_0+\langle kt\p_t+kx\p_x+my\p_y+2\p_u \rangle $ \\[0.5ex]
\refstepcounter{tbn}\label{caseFuGemuH0}\thetbn.\vspacebefore &  $u$ & $ge^{mu}$
         & $A^{\cap}_0+\langle my\p_y+2\p_u \rangle $ \\[0.5ex]
\refstepcounter{tbn}\label{caseFukGlnuH0}\thetbn.\vspacebefore &  $u^k$ & $g\ln u$
         & $A^{\cap}_0+\langle (k+2)t\p_t+kx\p_x+2u\p_u \rangle $ \\[0.5ex]
\refstepcounter{tbn}\label{caseFlnuGlnuH0}\thetbn.\vspacebefore &  $\ln u$ & $g\ln u$
         & $A^{\cap}_0+\langle t\p_t+u\p_u \rangle $ \\[0.5ex]
\refstepcounter{tbn}\label{caseFekuGuH0}\thetbn.\vspacebefore & $e^{ku}$ & $gu$
         & $A^{\cap}_0+\langle kt\p_t+kx\p_x+2\p_u \rangle $ \\[0.5ex]
\refstepcounter{tbn}\label{caseFuGu-13H0}\thetbn.\vspacebefore &  $u$ & $gu^{-1/3}$
         & $A^{\cap}_0+\langle {2y}\p_y-3u\sin{2y}\p_u,\, y^2\p_y-3yu\p_u \rangle $ \\[0.5ex]
\refstepcounter{tbn}\label{caseFuGuH0}\thetbn.\vspacebefore &  $u$ & $gu$
         & $A^{\cap}_0+\langle u\p_u,\, \varphi\p_u \rangle $ \\[0.5ex]
\hline
\end{tabular}
\end{center}
{\footnotesize
Here $g=\pm1$, $\varphi=\varphi(t,x,y)$ is an arbitrary solution of $\varphi_t+\varphi_{xxx}+g\varphi_{xyy}=0$.
}

\section{Exact solutions}\label{SectionExactSol}

In this section we construct exact solutions for some special forms of the GZK equations~\eqref{eq2DimDispZKEq} by means of the
classical Lie--Ovsiannikov algorithm. We first give a detailed classification of optimal systems of subalgebras of the Lie algebra of the equations
under consideration. Then we reduce the (2+1)-dimensional equations to (1+1)-dimensional equations or ordinary differentiate equation or algebraic equation
according to the subalgebras system and the corresponding ansatz.
Solving the reduced equations, we can finally obtain exact invariant solutions of the original equation.
Below, we illustrate this algorithm by two examples: the modified and classical Zakharov--Kuznetsov equations.

\subsection{Modified Zakharov--Kuznetsov equation}

Consider the modified Zakharov--Kuznetsov equation of form
\begin{equation}\label{eq2DimZKEqMod}
u_t+u_{xxx}+gu_{xyy}+hu^2u_x=0.
\end{equation}
It is shown above that equation~\eqref{eq2DimZKEqMod} is invariant with respect to a realization of
four-dimensional symmetry algebra $A^{\frac13,\frac13}_{4,5}$ spanned by the following operators
\[
v_1=\p_t,\quad v_2=\p_x,\quad\ v_3=\p_y,\quad v_4=t\p_t+\frac13x\p_x+\frac13y\p_y-\frac13u\p_u.
\]

The commutation relations of the symmetry operators are given by
\[ %$\displaystyle
\begin{array}{|c|cccc|}
\hline
[v_i,v_j] & v_1 &v_2&v_3& v_4\\
\hline
v_1&        0   & 0    &0     &v_1\\
v_2&        0   & 0    &0     &\frac13v_2\\[1ex]
v_3&        0   & 0    &0     &\frac13v_3\\[1ex]
v_4&    -v_1   & -\frac13v_2 &-\frac13v_3  &0\\
\hline
\end{array}
\] %$
Since $L_1$ has zero center, we can directly apply Ovsiannikov's method of classification of subalgebras~\cite{Ovsiannikov1982}.
Namely, construction of optimal system of one-dimensional subalgebras we start from taking a non-zero vector
\[
a_4v_4+a_3v_3+a_2v_2+a_1v_1
\]
and considering its image under the action of adjoint representations adduced in the following table.
\[
\begin{array}{|c|cccc|}
\hline
\mbox{Ad} & v_1 &v_2&v_3& v_4\\
\hline
v_1&                 v_1 &                v_2 &v_3                 &v_4-\epsilon v_1\\
v_2&                 v_1 &                v_2 &v_3                 &v_4-\frac13\epsilon v_2\\
v_3&                 v_1 &                v_2 &v_3                 &v_4-\frac13\epsilon v_3\\
v_4& e^{\epsilon}v_1 & e^{\frac13\epsilon}v_2 &e^{\frac13\epsilon}v_3  &v_4\\
\hline
\end{array}
\]

\begin{note}
Without going to the theoretical details we recall that the adjoint representation of a Lie group $G$ on its Lie algebra $L$
can be reconstructed from the infinitesimal adjoint action $\mbox{ad}L$ on itself by means of the following formula
\[
\mbox{Ad}(\exp(\epsilon v))w=\sum_{n=0}^{\infty}\frac{\epsilon^n}{n!}(ad v)^n(w)=w-\epsilon[v,w]+\frac{\epsilon^2}2[v,[v,w]]-\cdots.
\]
Here $v$ and $w$ are elements of the Lie algebra $L$,
$\exp(\epsilon v)$ is the one-parameter subgroup of~$G$ corresponding to the infinitesimal $v$.
For more details we refer the reader to~\cite{Olver1986}.
\end{note}

Then we try to choose the values of parameters in the adjoint actions
in order to simplify possible forms of the class of subalgebras that our vector belongs to.
Different possibilities arising under this procedure give us the classes of inequivalent one-dimensional subalgebras.
In such a way we find an optimal system of 1-dimensional subalgebras:
\[
\langle v_1+\varepsilon v_2+\varepsilon v_3 \rangle,\quad \langle v_2 \rangle,\quad \langle v_3+av_2 \rangle,\quad \langle v_4 \rangle,\quad
\langle v_1 \rangle.
\]
Here and below $\varepsilon=0,\pm1$, $a,b=\const$.

An excellent detailed explanation and examples of classification of one-dimensional subalgebras of Lie algebras can be found in
the textbooks~\cite{Olver1986,Ovsiannikov1982}.
(Inequivalent subalgebras of all 2-, 3- and 4-dimensional Lie algebras were classified in~\cite{Patera&Winternitz1977}.)

Constructing optimal system of two-dimensional subalgebras, we can suppose immediately that one of the basis vectors
of two-dimensional subalgebra is taken from the previously obtained optimal system of one-dimensional subalgebras.
Then we try to choose the parameters in the adjoint actions
in order to simplify possible forms of the second basis and not to ``spoil'' the first one.
It is possible that some of the basis vectors of the already classified one-dimensional subalgebras do not belong to any of the two-dimensional subalgebras.

After construction of all two-dimensional subalgebras for all basis vectors of the optimal system of one-dimensional subalgebras,
we have to consider the action of inner automorphisms to order and simplify them, similarly to what we have already done for one-dimensional subalgebras.
As a result we get an optimal system of two-dimensional subalgebras of the Lie symmetry algebra:
\begin{gather*}
\langle v_1, v_2 \rangle,\quad \langle v_1, v_3+av_2 \rangle,\quad
\langle v_2, v_3 \rangle \quad \langle v_3, v_1+\varepsilon v_2 \rangle,\quad
\langle v_1+\varepsilon v_3, v_2+av_3 \rangle,\\
\langle v_1, v_4 \rangle, \quad\langle v_4, v_3+av_2 \rangle, \quad\langle v_4, v_2 \rangle.
\end{gather*}

An optimal system of three-dimensional subalgebras can be constructed by means of extension of the two-dimensional subalgebras completely similarly
to the above extension of the one-dimensional subalgebras to the two-dimensional ones.
Note, that there exist three-dimensional algebras
(all isomorphic to $\langle e_1, e_2, e_3\rangle$, where $[e_1,e_2]=e_3$,  $[e_2,e_3]=e_1$,  $[e_1,e_3]=-e_2$)
that do not contain two-dimensional subalgebras.
However, the symmetry algebra $L_1$ is solvable, therefore, all its three-dimensional subalgebras contain two-dimensional subalgebras
and we can directly extend the above classification to construction of the three-dimensional subalgebras.

In such a way we get an optimal system of 3-dimensional subalgebras of $L_1$ that consists of
\[
\langle v_1,v_2,v_3 \rangle,\quad \langle v_2, v_3, v_4 \rangle,\quad \langle  v_1, v_4, v_3+av_2\rangle,\quad \langle  v_1, v_2, v_4\rangle.
\]

As one can easily check, the obtained optimal systems coincide with those adduced in~\cite{Patera&Winternitz1977}.

\medskip
Now we consider reductions of equation~\eqref{eq2DimZKEqMod} with respect to the obtained inequivalent subalgebras.

\bigskip

\centerline{\bf Reductions with respect to 1-dimensional subalgebras}

\medskip

We try to reduce equation~\eqref{eq2DimZKEqMod} with respect to a subalgebra generated by a Lie symmetry of the form
\begin{gather*}
\tau(t,x,y,u)\p_t+\xi(t,x,y,u)\p_x+\zeta(t,x,y,u)\p_y+\eta(t,x,y,u)\p_u ,
\end{gather*}
where $\tau$, $\xi$, $\zeta$ and $\eta$ are known functions. In order to derive the desired similarity reductions,
we need to solve a partial differential equation of form
\[
\tau u_t+\xi u_x+\zeta u_y=\eta .
\]
The solution of this equation contains three independent integrals which provide the general solution in the form
\begin{equation}\label{AnsTo2Dim}
u=\mu(t,x,y,w(p,z)),\quad p=p(t,x,y),~ z=z(t,x,y),
\end{equation}
where $\mu$ is known function and $v$, $p$, $z$ are arbitrary functions in their arguments.
Formula~\eqref{AnsTo2Dim} defines the Ansatz that maps~\eqref{eq2DimZKEqMod} into
a partial differential equation with two independent variables $p$ and $z$, and $w$ being
the dependent variable.

Below in case 1 of the reductions with respect to 1-dimensional subalgebras we consider this procedure in details.
In the rest of the cases we adduce only the Ansatz, new independent variables, the reduced equations and,
in some cases, corresponding invariant solutions.

\medskip
\noindent
1.
$\langle v_1+\varepsilon v_2+\varepsilon v_3 \rangle$:
We try to construct invariant reduction with respect to the operator $Q=\p_t+\varepsilon \p_x+a\p_y$.
In the framework of the above procedure we look for the first integrals of the equation $u_t+\varepsilon u_x+au_y=0$.
They can be chosen as $I_1=u$, $I_2=y-at$ and $I_3=x-\varepsilon t$.
Then, the general solution of the above first-order equation has the form $F(I_1,I_2,I_3)=0$, where $F$ is an arbitrary function of its variables.
Expressing now $u$ from the above equation, we get the Ansatz  $u=w(p,z)$, $p=y-at$, $z=x-\varepsilon t$ for the reduction.
Substituting this Ansatz to the initial equation yields the reduced equation with two independent variables that looks like
$-aw_p-\varepsilon w_z+w_{zzz}+gw_{zpp}+hw^2w_z=0$.

Solving this reduced equations and using the above-mentioned ansatz, we can obtain exact solutions of the
modified ZK equation \eqref{eq2DimZKEqMod}.
We succeeded to find some partial solutions of the reduced equation
that provide us with the following solutions of the modified ZK equation~\eqref{eq2DimZKEqMod}.
%\marginpar{\footnotesize check!!! no correspondence between $\tan$ and $\tanh$!!!}
\begin{gather*}
u=A\csc\phi,
\quad
u=A\sec\phi,
\quad
u=\frac A{\sqrt2}\tanh\frac{\phi}{\sqrt2},
\quad
u=\frac A{\sqrt2}\coth\frac{\phi}{\sqrt2},\\
u=A\sech\phi,\quad
u=A\csch\phi,\quad
u=\frac A{\sqrt2}\tan\frac{\phi}{\sqrt2},
\quad
u=\frac A{\sqrt2}\cot\frac{\phi}{\sqrt2},
\end{gather*}
where $A=\sqrt{\left|\frac{6(ac_1+\varepsilon c_2)}{hc_2}\right|}$,
$\phi=\sqrt{\left|\frac{ac_1+\varepsilon c_2}{(c_2^3+gc_2c_1^2)}\right|}(-(ac_1+\varepsilon c_2)t+c_2x+c_1y)$,
$c_1, c_2$ are arbitrary constants which satisfy the constrains $\frac{6(ac_1+\varepsilon c_2)}{hc_2}>0,~\frac{ac_1+\varepsilon c_2}{c_2^3+gc_2c_1^2}<0$
for the first five solutions and $\frac{6(ac_1+\varepsilon c_2)}{hc_2}<0,~\frac{ac_1+\varepsilon c_2}{c_2^3+gc_2c_1^2}>0$ for the last three solutions.
It should be noted that the third to the sixth solutions are solitary wave solutions, while the other four solutions are periodic
wave solutions.

\medskip
\noindent
2.
$\langle v_2 \rangle$: $u=w(t,y)$, $w_t=0$ that gives trivial solution $u=u(y)$.

\medskip
\noindent
3.
$\langle v_3+av_2 \rangle$: $u=w(p,z)$, $p=t$, $z=x-ay$,
 $w_p+w_{zzz}+a^2gw_{zzz}+hw^2w_z=0$.
 This reduced equation is the well-known modified KdV equations,
 and some of its known invariant solutions provide us with the follownig solutions
  for the modified ZK equation \eqref{eq2DimZKEqMod}:
%  \marginpar{\footnotesize please put similarly to page 17}
\begin{gather*}
u=A\csc\phi,
\quad
u=A\sec\phi,
\quad
u=\frac A{\sqrt2}\tanh\frac{\phi}{\sqrt2},
\quad
u=\frac A{\sqrt2}\coth\frac{\phi}{\sqrt2},\\
u=A\sech\phi,\quad
u=A\csch\phi,\quad
u=\frac A{\sqrt2}\tan\frac{\phi}{\sqrt2},
\quad
u=\frac A{\sqrt2}\cot\frac{\phi}{\sqrt2},
\end{gather*}
where $A=\sqrt{\left|\frac{6c_1}{hc_2}\right|}$,
$\phi=\sqrt{\left|\frac{c_1}{c_2^3+a^2gc_2^3}\right|}(c_1t+c_2x-ac_2y)$,
$c_1, c_2$ are arbitrary constants which satisfy the constrains $\frac{6c_1}{hc_2}>0,~\frac{c_1}{c_2^3+a^2gc_2^3}<0$
for the first five solutions and $\frac{6c_1}{hc_2}<0,~\frac{c_1}{c_2^3+a^2gc_2^3}>0$ for the last three solutions.

\medskip
\noindent
4.
$\langle v_4 \rangle$: $u=t^{-1/3}w(p,z)$, $p=t^{-1/3}y$, $z=t^{-1/3}x$,
 $-\frac13w-\frac13pw_p-\frac13zw_z+w_{zzz}+gw_{ppz}+hw^2w_z=0$.

\medskip
\noindent
5.
$\langle v_1 \rangle$: $u=w(x,y)$,
 $w_{xxx}+gw_{xyy}+hw^2w_x=0$. Solving this reduced equation, we can obtain stationary solutions of the modified ZK equation~\eqref{eq2DimZKEqMod}
 of form:
 \[
 u=w(kx+ly),
 \]
where $w$ is implicitly determined by
\[
\int \sqrt{\frac{6(k^2+gl^2)}{6c_2w+6c_1-hw^4}}dw=\pm (kx+ly)+c_0,
\]
where $c_2,~c_1,~c_0, ~k, ~l$ are arbitrary constants.

\bigskip

\centerline{\bf Reductions with respect to 2-dimensional subalgebras}

\medskip
\noindent
1.
$\langle v_2, v_1 \rangle$: leads to the solution $u=v(y)$

\medskip
\noindent
2.
$\langle v_1, v_3+av_2 \rangle$: $u=v(x-ay)$, $(1+g^2)v'''+hv^2v'=0$.
Solving this reduced equation, we obtain stationary solutions of the modified ZK equation \eqref{eq2DimZKEqMod} in implicit form:
\[
u=v(\xi),\quad \int \sqrt{\frac{6(1+g^2)}{12c_2v+12c_1-hv^4}}dv=\pm \xi+c_0,\quad \xi=x-ay,
\]
where $c_2,~c_1,~c_0$ are arbitrary constants.

\medskip
\noindent
3.
$\langle v_2, v_3 \rangle$: leads to the solution $u=\const$.

\medskip
\noindent
4.
$\langle v_3, v_1+\varepsilon v_2 \rangle$: $u=v(x-\varepsilon t)$, $v'''-\varepsilon v'+hv^2v'=0$.
Solving this reduced equation, we get a $y$-independent general solutions of the modified ZK equation \eqref{eq2DimZKEqMod} in implicit form:
\[
u=v(\xi),\quad \int \frac{1}{\sqrt{\varepsilon v^2+2c_2v+2c_1-\frac{1}{6}hv^4}}dv=\pm \xi+c_0,\quad \xi=x-\varepsilon t,
\]
where $c_2,~c_1,~c_0$ are arbitrary constants.

\medskip
\noindent
5.
$\langle v_1+\varepsilon v_3, v_2+av_3 \rangle$: $u=v(y-ax-\varepsilon t)$, $(a^3+ga)v'''+\varepsilon v'+ahv^2v'=0$.
Solving this reduced equation, we find solutions of the modified ZK equation \eqref{eq2DimZKEqMod} in implicit form:
\[
u=v(\xi),\quad \int \sqrt{\frac{a(a^2+g)}{2c_2v+2c_1-\varepsilon v^2-\frac{1}{6}ahv^4}}dv=\pm \xi+c_0,\quad \xi=y-ax-\varepsilon t,
\]
where $c_2,~c_1,~c_0$ are arbitrary constants.

\medskip
\noindent
6.
$\langle v_1, v_4 \rangle$: $u=\frac{v(\omega)}y $, $\omega=\frac xy $, $(1+g\omega^2)v'''+6g\omega v''+6gv'+hv^2v' =0$. Integrating the equation one time, we obtain
\[
[(1+gw^2)v]''+\frac{1}{3}hv^3=c_0.
\]
Solving this equation, we obtain a stationary solution of the modified ZK equation \eqref{eq2DimZKEqMod}:
\[
u(x,y)=\frac{c_2}{y\sqrt{1+g(x/y)^2}}m\sn[m(\frac{\sqrt{36g+6h}\arctan(\sqrt{g}(x/y))}{6\sqrt{g}}+c_1),
\frac{c_2\sqrt{-6gh-h^2}}{6g+h}],
\]
where $m=\sqrt{\frac{6g}{6g+h-hc_2^2}}$, $c_1,~c_2$ are arbitrary constants and $\sn(\cdot,\cdot)$ is Jacobi elliptic sine function.

\medskip
\noindent
7.
$\langle v_4, v_3+av_2 \rangle$: $u=t^{-1/3}v(\omega)$, $\omega=t^{-1/3}(x-ay)$, $(1+a^2g)v'''-\frac13\omega v'-\frac13v+hv^2v'=0$. Solving this equation, we obtain stationary solution of the modified ZK equation \eqref{eq2DimZKEqMod}:
\[
u=\pm \sqrt{-\frac{6(1+a^2g)}{h}}\frac{1}{x-ay},
\]
where $\frac{6(1+a^2g)}{h}<0$.

\medskip
\noindent
8.
$\langle v_4, v_2 \rangle$:  leads to the solution of form $u=v(y)$.

\bigskip

\centerline{\bf Reductions with respect to 3-dimensional subalgebras}
\medskip

Since equation~\eqref{eq2DimZKEqMod} is 3-dimensional, its reductions with respect to 3-dimensional subalgebras~$L_1$ lead to algebraic equations.
From invariance of~\eqref{eq2DimZKEqMod} with respect to $\langle  v_1, v_4, v_3+av_2\rangle$ we obtain that
\[
u=\frac c{x-ay}.
\]
Substituting this expression to~\eqref{eq2DimZKEqMod} we get an algebraic equation on $c$ of form
$c(6+6a^2+hc^2)=0$. Its solutions are
\[
c=0,\quad c=\pm\frac{\sqrt{-h(a^2+1)}}h.
\]

It is not difficult to show that all other reductions with respect to 3-dimensional subalgebras lead, at most, to the trivial solution
\[
u=u(y).
\]

\subsection{Zakharov--Kuznetsov equation}

In this subsection perform reductions of the Zakharov--Kuznetsov equation
\begin{equation}\label{eq2DimZKEqClass}
u_t+u_{xxx}+gu_{xyy}+huu_x=0.
\end{equation}
with respect to 1- and 2-dimensional subalgebras of its maximal Lie symmetry algebra
\[
\langle
v_1=\p_t,\ v_2=\p_x,\ v_3=\p_y,\ v_4=ht\p_x+\p_u,\ v_5=3t\p_t+x\p_x+y\p_y-2u\p_u
 \rangle.
\]

As the computations are very similar to the ones from the previous subsection, we skip all technical details and summarize the results only.
(Some of the reduced with respect to 1-dimensional subalgebras equations together with their solutions can be found also in~\cite{Moussa2001}.)
The table of commutation relations is
\[
\begin{array}{|c|ccccc|}
\hline
[v_i,v_j] & v_1 &v_2&v_3& v_4 &v_5\\
\hline
v_1&  0    & 0     &0      & hv_2  &3v_1\\
v_2&  0    & 0     & 0     &0      &v_2\\
v_3&  0    & 0     & 0     & 0     & v_3\\
v_4& -hv_2 & 0     & 0     & 0     & -2v_4 \\
v_5& -3v_1 & -v_2  & -v_3  & 2v_4  & 0\\
\hline
\end{array}
\]
Then, the corresponding adjoint representations look like
\[
\begin{array}{|c|ccccc|}
\hline
\mbox{Ad} & v_1 &v_2&v_3& v_4&v_5\\
\hline
v_1&  v_1                  & v_2                &v_3                  & v_4-h\epsilon v_2  & v_5-3\epsilon v_1\\
v_2&  v_1                  & v_2                & v_3                 & v_4                   & v_5-\epsilon v_2\\
v_3&  v_1                  & v_2                & v_3                 & v_4                   & v_5-\epsilon v_3\\
v_4 & v_1+h\epsilon v_2 & v_2                & v_3                 & v_4                   & v_5+2\epsilon v_4 \\
v_5 & e^{3\epsilon}v_1  & e^{\epsilon}v_2 & e^{\epsilon}v_3  & e^{-2\epsilon}v_4  & v_5\\
\hline
\end{array}
\]
An optimal system of 1-dimensional subalgebras can be chosen as
\[
\langle v_1 \rangle,\quad \langle v_2 \rangle,\quad \langle v_3+a_2v_2 \rangle, \quad \langle v_3+\varepsilon_1v_1 \rangle,
\quad \langle v_4+a_3v_3+a_1v_1 \rangle, \quad \langle v_5 \rangle.
\]
Here and below $\varepsilon_i=0,\pm1$, $a_i,b_i=\const$.

An optimal system of 2-dimensional subalgebras is
\begin{gather*}
\langle v_2, v_1 \rangle,\quad \langle v_3, v_2 \rangle,\quad \langle v_3+\varepsilon_2v_2, v_1 \rangle,\quad
\langle v_4+\varepsilon_3v_3, v_1 \rangle, \quad \langle v_4+a_3v_3, v_2 \rangle, \\
 \langle v_4+a_1v_1, v_3 \rangle, \quad
\langle v_5, v_2 \rangle,\quad \langle v_5, v_1 \rangle,\quad \langle v_5, v_3-b_2v_2 \rangle,\quad \langle v_5, v_4 \rangle
\end{gather*}

\centerline{\bf Reductions with respect to 1-dimensional subalgebras}

\medskip
\noindent
1.
$\langle v_1 \rangle$: $u=w(x,y)$, $w_{xxx}+gw_{xyy}+hww_x=0$.
An invariant solution arising from this case is
\[
w(x,y)=\frac{8}{h}(c_2^2+gc_3^2)-\frac{12}{h}(c_2^2+gc_3^2)\tanh^2 (c_1+c_2x+c_3y),
\]
where $c_1,~c_2,~c_3$ are arbitrary constants.

\medskip
\noindent
2.
$\langle v_2\rangle$: $u=w(t,y)$, $w_t=0$, $u=w(y)$

\medskip
\noindent
3.
$\langle v_3+a_2v_2 \rangle$: $u=w(t,z)$, $z=x-a_2y$, $w_t+(1+a_2^2g)w_{zzz}+hww_z=0$.
This equation is the famous KdV equation.
Using some of its invariant solutions we can easily construct some exact solutions for the ZK equation \eqref{eq2DimZKEqClass}:
%\marginpar{\footnotesize please put similarly to page 17}

\begin{gather*}
u=A\csc^2\phi,
\quad
u=A\sec^2\phi,
\quad
u=-\frac A{3}(1-3\tanh^2\phi),
\quad
u=-\frac A{3}(1-3\coth^2\phi),\\
u=A\sech^2\phi,\quad
u=-A\csch^2\phi,\quad
u=-\frac A{3}(1+3\tan^2\phi),
\quad
u=-\frac A{3}(1+3\cot^2\phi),
\end{gather*}
where $A=-\frac{3c_1}{hc_2}$,
$\phi=\frac{1}{2}\sqrt{\left|\frac{c_1}{c_2^3+a_2^2gc_2^3}\right|}(c_1t+c_2x-a_2c_2y)$,
$c_1, c_2$ are arbitrary constants which satisfy the constrains $\frac{c_1}{c_2^3+a_2^2gc_2^3}>0$ for the first
five solutions and $\frac{c_1}{c_2^3+a_2^2gc_2^3}<0$ for the last three solutions.
It should be noted that the third to the sixth solutions are solitary wave solutions, while the other four solutions are periodic
wave solutions.

\medskip
\noindent
4.
$\langle v_3+\epsilon_1v_1 \rangle$: $u=w(p,x)$, $p=t-\varepsilon_1y$, $w_p+w_{xxx}+\varepsilon_1^2gw_{ppx}+hww_x=0$.
Solving this reduced equation, we can obtain the following exact solutions of the ZK equation \eqref{eq2DimZKEqClass}:
%\marginpar{\footnotesize please put similarly to page 17}
\begin{gather*}
u=A\csc^2\phi,
\quad
u=A\sec^2\phi,
\quad
u=-\frac A{3}(1-3\tanh^2\phi),
\quad
u=-\frac A{3}(1-3\coth^2\phi),\\
u=A\sech^2\phi,\quad
u=-A\csch^2\phi,\quad
u=-\frac A{3}(1+3\tan^2\phi),
\quad
u=-\frac A{3}(1+3\cot^2\phi),
\end{gather*}
where $A=-\frac{3c_1}{hc_2}$,
$\phi=\frac{1}{2}\sqrt{\left|\frac{c_1}{c_2^3+\varepsilon_1^2gc_1^2c_2}\right|}(c_1t+c_2x-a_2c_2y)$,
$c_1, c_2$ are arbitrary constants which satisfy the constrains $\frac{c_1}{c_2^3+\varepsilon_1^2gc_1^2c_2}>0$ for the first five solutions and $\frac{c_1}{c_2^3+\varepsilon_1^2gc_1^2c_2}<0$ for the last three solutions.

\medskip
\noindent
5.
$\langle v_4+a_3v_3+a_1v_1 \rangle$: if $a_1=0$, then $u=\frac x{ht}+w(t,z)$, $z=\frac{a_3x}{ht}-y$,
$h^3t^3w_t+w_{zzz}(a_3^3+ga_3h^2t^2)+h^3t^2w+h^3t^2a_3ww_z=0$.
If $a_1\ne0$, $u=\frac t{a_1}+w(p,z)$, $z=x-\frac{ht^2}{2a_1}$, $p=a_1y-a_3t$,
$\frac1{a_1}-a_3w_p+w_{zzz}+ga_1^2w_{zpp}+hww_z=0$.

\medskip
\noindent
6.
$\langle v_5 \rangle$: $u=t^{-2/3}w(p,z)$, $p=t^{-1/3}y$, $z=t^{-1/3}x$, $-2w-pw_p-zw_z+3w_{zzz}+3gw_{zpp}+3hww_z=0$.

\bigskip

\centerline{\bf Reductions with respect to 2-dimensional subalgebras}

\medskip
\noindent
1.
$\langle v_2, v_1 \rangle$: $u=v(y)$ is a solution

\medskip
\noindent
2.
$\langle v_3, v_2 \rangle$: $u=v(t)$, $v_t=0$, therefore, $u=\const$.

\medskip
\noindent
3.
$\langle v_3+\varepsilon_2v_2, v_1 \rangle$: $u=v(x-\varepsilon_2y)$, $(1+g\varepsilon_2^2)v'''+hvv'=0$.
Solving this reduced equation, we obtain stationary solution of the ZK equation \eqref{eq2DimZKEqClass} in implicit form:
\[
u=v(\xi),\quad \int \sqrt{\frac{(1+g\varepsilon_2^2)}{2c_2v+2c_1-\frac{1}{3}hv^3}}dv=\pm \xi+c_0,\quad \xi=x-\varepsilon_2y,
\]
where $c_2,~c_1,~c_0$ are arbitrary constants.

\medskip
\noindent
4.
$\langle v_4+a_3v_3, v_2 \rangle$: if $a_3=0$ then there is no reduction,
if $a_3\ne0$, then $u=\frac y{a_3}+v(t)$, $v_t=0$ that gives $u=\frac y{a_3}+c$, where $c$ is an arbitrary constant.

\medskip
\noindent
5.
$\langle v_4+a_1v_1, v_3 \rangle$: if $a_1=0$, then $u=\frac x{ht}+v(t)$, $v'+\frac vt=0$ that gives $u=\frac x{ht}+\frac{c}{t}$,
where $c$ is an arbitrary constant.
If $a_1\ne0$, then $u=\frac t{a_1}+v(x-\frac h{2a_1}t^2)$, $v'''+hvv'+\frac1{a_1}=0$.
Integrating this equation one time, we obtain $v''+\frac{1}{2}hv^2+\frac1{a_1}\xi=c$, where $\xi=x-\frac h{2a_1}t^2$.

\medskip
\noindent
6.
$\langle v_5, v_2 \rangle$: $u=t^{-2/3}v(\omega)$, $\omega=t^{-1/3}y$, $2v+\omega v'=0$.
Solving this equation, we obtain solution of the ZK equation \eqref{eq2DimZKEqClass} of form $u=\frac{c_0}{y^2}$.

\medskip
\noindent
7.
$\langle v_5, v_1 \rangle$: $u=\frac1{y^2}v(\omega)$, $\omega=\frac xy$, $(1+g\omega^2)v'''+12gv'+8g\omega v''+hvv'=0$. Integrating the equation one time, we obtain
\[
(1+gw^2)v''+6g(wv)'+\frac{1}{2}hv^2=c_0.
\]
Solving this reduced equation with $c_0=0$, we obtain stationary solution of the ZK equation~\eqref{eq2DimZKEqClass} in implicit form:
\[
u=\frac{1}{y^2}v(\omega),\quad \frac{1}{\sqrt{g}}\arctan(\sqrt{g}\omega)\pm 3 \int \frac{1}{\sqrt{9c_1-36gZ^2-3hZ^3}}dZ-c_2=0,\quad \omega=\frac xy,
\]
where $Z=v+gv\omega^2,~c_2,~c_1$ are arbitrary constants.

\medskip
\noindent
8.
$\langle v_5, v_3-b_2v_2 \rangle$: $u=t^{-2/3}v(\omega)$, $\omega=t^{-1/3}(x+b_2y)$, $3(1+gb_2^2)v'''-2v-\omega v'+3hvv'=0$. This reduced equation has a solution $v=\frac{1}{h}\omega$,
thus we obtain a rational solution of the ZK equation~\eqref{eq2DimZKEqClass}: $u=\frac{x+b_2 y}{ht}$.

\medskip
\noindent
9.
$\langle v_5, v_4 \rangle$: $u=\frac x{ht}+t^{-2/3}v(\omega)$, $\omega=t^{-1/3}y$, $\omega v'-v=0$ that gives $u=\frac x{ht}+\frac{cy}{t}$, where $c$ is an arbitrary constant.

\begin{note}
In both cases of the modified Zakharov--Kuznetsov equation and of the classical Zakharov--Kuznetsov equation, we do not solve all the reduced equations.
One of the possible ways of constructing  their solutions is, again, group analysis.
More precisely, one can  derive Lie symmetries for the reduced partial differential equations with the
ultimate goal to construct similarity reductions that transform these equations into ordinary
differential equations or algebraic equations that are easier to solve.
\end{note}

\section{Conclusion and Discussion}\label{SectionOnConclusion}
In summary, we have described an algorithmic framework for group classification of (systems of) partial differential equations. More precisely,
we have revised the classical Lie--Ovsiannikov method of construction of symmetries (of system) of differential equations in more details
and written down the precise formulation of the algorithm of group classification of a class (of systems) of differential equations.
We also described a systematic way of finding all possible invariant solutions of (of systems) of differential equations.
All the theory and algorithms were illustrated by the running example of the GZK equations \eqref{eq2DimDispZKEq}.
Specifically, we performed a complete group classification of
the class of GZK equations~\eqref{eq2DimDispZKEq} by using the
equivalence transformations and the compatibility method.
The main results of classification are collected in Tables~\ref{TableGrCl2DimDispZKEq}--\ref{TableGrCl2DimDispZKEqH0} where we list all inequivalent cases of extensions with the corresponding Lie invariance algebras.
It is shown that the symmetry algebras of invariant models of form \eqref{eq2DimDispZKEq}, are at most
six-dimensional. For the classical Zakharov--Kuznetsov \eqref{eq2DimZKEqClass} and the modified Zakharov--Kuznetsov \eqref{eq2DimZKEqMod} we construct optimal systems of inequivalent subalgebras, corresponding Lie ans\"{a}tze and invariant solutions.

The present paper should be an inspiration for further
investigations of other properties of class~\eqref{eq2DimDispZKEq}. For example,
one can classify the nonclassical (conditional) symmetries.
Furthermore, one can perform conservation law classification, then to use these results to
systematical  calculation of nonlocal symmetries, higher order local and potential conservation
laws, nonclassical potential solutions and linearizations, etc. Motivated by Craddock and Dooley's work on
the equivalence of Lie symmetries and group representations \cite{Craddock&Dooley2010}, we can also consider the global action of the Lie symmetries of the GZK equations \eqref{eq2DimDispZKEq}. We will investigate these subjects elsewhere. The algorithmic framework for group analysis of differential equations presented in this paper can also be applied
to other nonlinear PDEs in mathematical physics.

\subsection*{Acknowledgements}
Research of D.J.H. was supported by the special Postdoctoral Science
Foundation of China under Grant No. 201104247. Research of N.M.I. was partially supported by Cyprus Research Promotion Foundation
(project number $\Pi$PO$\Sigma$E$\Lambda$KY$\Sigma$H/$\Pi$PONE/0308/01).

\end{document}